\documentclass[secnumarabic,preprint]{revtex4-1}
\usepackage{amssymb,amsbsy,amsmath}
\usepackage[english]{babel}

\newcommand{\ha}{\mbox{\small$\frac{1}{2}$}}

\newcommand{\im}{\mathrm{i}\,}
\newcommand{\hi}{\mbox{\small$\frac{\mathrm{i}}{2}$}}
\newcommand{\na}{{\bar a}}
\newcommand{\lab}[1]{\label{#1}}
\newcommand{\re}[1]{(\ref{#1})}
\newcommand{\nn}{\nonumber}
\newcommand{\by}[1]{\stackrel{\mathrm{#1}}{=\!=}}
\newcommand{\B}[1]{\boldsymbol{#1}}
\newcommand{\s}[1]{\mathsf{#1}}
\newcommand{\sOm}{\mathsf{\Omega}}
\newcommand{\BOm}{\boldsymbol{\Omega}}

\newcommand{\BUp}{\boldsymbol{\mathit\Upsilon}}
\newcommand{\vez}{\rule{0pt}{2.3ex}\smash{\stackrel{\parallel}{\rule{0pt}{1.2ex}0}}}
\newcommand{\inta}{\hspace*{-.3em}\int \hspace*{-.4em}}
\newcommand{\intab}{\hspace*{-.3em}\int \hspace*{-.6em}\int \hspace{-.4em}}
\newcommand{\ints}{\lefteqn{\hspace*{-.2em}\int \hspace*{-.8em}\int
\hspace{-.3em}}\hspace{.1em}\diagdown\,}
\newcommand{\D}[2]{{\rm d}^{#1}{#2}\,}
\newcommand{\cc}[1]{\lefteqn{\stackrel{*}{\phantom{#1}}}#1}
\newcommand{\La}{{\cal L}}
\begin{document}


\title{Quantization of almost-circular orbits in the Fokker action formalism.
General scheme.}
\author{Askold Duviryak}
\email[]{duviryak@icmp.lviv.ua}
\affiliation{Department for Computer
Simulations of Many-Particle Systems, Institute for Condensed Matter
Physics of NAS of Ukraine, Lviv, UA-79011, Ukraine}

\date{\today}

\begin{abstract}
General two-particle system is considered within the formalism of
Fokker-type action integrals.
It is assumed that the system is invariant with respect to
the Aristotle group which is a common subgroup of the Galileo and Poincar\'e
groups.
It is shown that equations of motion of such system admit circular orbit solutions.
The dynamics of perturbations of these solutions is studied.
It is described by means of a linear homogeneous set of time-nonlocal equations
and is analyzed in terms of eigenfrequencies and eigenmodes.
The Hamiltonian description of the system
is built in the almost circular orbit approximation.
The Aristotle-invariance of the system is exploit to
avoid a double count of degrees of freedom and to select
physical modes.
The quantization procedure and a construction of energy spectrum
of the system is proposed.
\end{abstract}


\maketitle


\section{Introduction}
\renewcommand{\theequation}{1.\arabic{equation}}
\setcounter{equation}{0}

Fokker-type action integrals \cite{Hav71,Ker72} represent an
approach to the relativistic particle dynamics which is alternative
or complementary (depending on a point of view) to field-theoretical
approaches. Known about for a century \cite{Sch03,Tet22,Fok29}, but
mainly owing to the Wheeler-Feynman electrodynamics
\cite{W-F45,W-F49}, this approach was generalized to other
field-type relativistic interactions including cases of higher-rank
tensor fields \cite{Ram83,Tre96,Tre11},  gravitation
\cite{Tur82,V-T86,H-N74,H-N95}, confining interactions
\cite{Riv84,Wei86,Duv99,Duv01,L-M12} etc \cite{L-M06}.

A variational problem based on the Fokker-type action describes a
dynamical system with time non-locality, i.e., it leads to
difference-differential or integral-differential equations of motion
for which the Cauchy problem is unsuitable. Consequently, the study
the phase space (i.e., a set of possible states), the construction
of the Hamiltonian description and quantization of such a system are
non-trivial tasks.

Serious effort was made to develop Hamiltonization procedure for the
Fokker-type action integrals. In general, this is attained by means
of reformulation of the problem into another but time-local form.
Here we just mention two such schemes.

The first is a formal expansion of the Fokker-type action into the
Lagrangian action with higher derivatives (of order up to infinity)
\cite{W-H72,G-T80}, with a subsequent use of a modified
Hamilton-Ostrogradsky formalism \cite{GKT87}. In the second scheme
developed by Llosa et al. \cite{JJLM87,L-V94} the variational
problem is reformulated into a static one for particle world lines
treated as temporally extended strings. In practice both schemes can
be realized approximately: in the first the quasi-relativistic
approximations \cite{GKT87} are used, for the second the
coupling-constant expansion method was developed \cite{JJLM89}. Thus
the resulting Hamiltonian description of an N-particle system is
built on the 6N-dimensional phase space, as in a non-relativistic or
free-particle case.

Among not numerous solutions to Fokker-type variational problems
studied in literature the class of two-particle circular orbit {\em
exact} solutions \cite{Sch63,A-B70,Deg71} is of particular interest.
These solutions include domain of essentially relativistic motion of
strongly coupled particles and thus they stand apart the field of
application of quasi-relativistic approximations and a coupling
constant expansion. In the case of the Wheeler-Feynman
electrodynamics equations for small deviations from circular orbit
were derived and studied \cite{A-B72}. The analysis revealed
bifurcation points in the highly relativistic domain of the phase
space where redundant (as to compare to a non-relativistic case)
unstable degrees of freedom get excited. This result was approved by
direct numerical \cite{KNU98,N-L01} and global analytical
\cite{Luc09} study of the Wheeler-Feynman two-body variational
problem. We do not discuss here a physical meaning of highly
relativistic unstable solutions mentioned above. But on the whole
the almost circular orbit (ACO) approximation turns out more
informative and thus appropriate in the highly relativistic domain
than the quasi-relativistic or weak coupling approximations.

Of physical interest is a study, in ACO approximation, of various
Fokker-type systems, especially those which may have relevance to
the relativistic bound state problems in the nuclear and hadronic
physics. In particular, some Fokker-type systems with confining
interaction may serve as relativistic potential model of mesons
\cite{Riv84,Wei86,Duv99,Duv01}. Thereupon the quantization procedure
of Fokker-type models which is based on ACO approximation scheme
should be developed. One can relay in this way on the Bohr
quantization of circular orbits \cite{Deg71} and a heuristic
suggestion \cite{Bay75} to use the Miller's quantum condition
\cite{Mil75} for WKB quantization of ACO in the Wheeler-Feynman
electrodynamics.

In this paper we propose a substantiate quantization recipe of a
two-body Fokker-type problem of general form in ACO approximation.
The recipe in based on an implicit Hamiltonian description of the
system which, in turn, is built by means of Llosa scheme
\cite{L-V94}. We suppose that a Fokker-type system is invariant with respect to
the Aristotle group which is a common subgroup of the Galilei- and
Poincar\'e groups. By this both non-reativistic as well as relativistic
systems are involved into consideration. The symmetry with respect to time
translations and space rotations results in the existence of
generalized Noether integrals of motions \cite{Her85}, i.e., the
energy and the total angular momentum. We show that, under rather
general condition, a system admits a circular particle motion with a
given constant angular velocity. It is taken as zero-order
approximation for non-circular motions.
In order to study a perturbation to circular
orbit solution we use a uniformly rotating reference frame where
circulating particles are motionless. Then ACO solution is
formulated in terms of small deviations from fixed (and presumably
equilibrium) particle positions. We obtain a time-nonlocal
action principle for these deviations and derive corresponding linear
homogeneous set of integral-differential equations of motion.
Fundamental set of solutions to these equations can be expressed in
terms of characteristic frequencies and amplitudes of generalized
normal modes. The amplitudes are shown to be canonical variables,
the frequencies are functions of the total angular momentum, and all
they constitute a correction to zero-order
circular-orbit energy. Then the quantization is trivial. In general,
it must be complemented by some selection rules for separation of
physical modes out from all variety of them. It is discussed in
details.

The paper is organized as follows. In Sec. II we apply the ACO
approximation method to a single-particle Lagrangian system which is
rotary-invariant and local in time. This section has a rather
methodological meaning since main points of the method are
demonstrated, and useful definitions and notations are introduced
there. In Sec. III the method is extended for a general
Galilei-invariant two-particle system. We consider the latter as a
time-local or slow-motion limit of a wide class of Fokker-type
two-particle systems examined in Sec. IV. Perturbations to
circular orbits are shown to be described by a linear set of
time-nonlocal equations of motion. Symmetry and dynamical
properties of this set is studied in various subsections of
the Sec. IV as well as in appendices. In particular, in Appendix D
the Hamiltonization and quantization of a linear
nonlocal system is discussed in detail.


\section{Rotary-invariant single-particle dynamics}
\renewcommand{\theequation}{2.\arabic{equation}}
\setcounter{equation}{0}

Let us consider a system of single particle which is invariant under
the time
translations $t\to t+\lambda$ and the space rotations $\B x\to\s R\B x$,
where $\lambda\in\Bbb R$, $\s R \in \mathrm{O(3)}$ and
$\B x\equiv\{x_i;\ i{=}1,2,3\}\in\Bbb E^3$.
The Lagrangian function $L(\B x,\dot{\B x})$,
satisfies the equality:
%
\begin{equation}\label{2.1}
L(\s{R}\B x,\s{R}\dot{\B x}) = L(\B x,\dot{\B x})
\end{equation}
and thus has the following structure:
%
\begin{equation}\label{2.2}
L(\B x,\dot{\B x}) = L(\B x^2,\,\B x\cdot\dot{\B x},\,\dot{\B x}^2)
\equiv L(\alpha, \beta, \gamma).
\end{equation}
Following the Noether theorem the energy $E$ and angular momentum $\B J$,
%
\begin{eqnarray}
&& E = \B x\cdot\B p - L, \label{2.3}\\
&& \B J = \B x\B\times\B p, \label{2.4}
\end{eqnarray}
are conserved; here $\B p = \partial L/\partial\dot{\B x}$ and
"$\B\times$" denotes a vector product.

The system is exactly integrable with standard methods. We consider
this example here in order to demonstrate the idea of the
approximate method applied below to a Fokker-type system.

\subsection{The description in a uniformly rotating
reference frame}

First of all we perform the coordinate transformation $\B x\mapsto \B z$
corresponding to transition to a uniformly rotating reference frame:
%
\begin{equation}\label{2.5}
\B x(t) = \s S(t)\B z(t), \qquad\mbox{with}\quad  \s S(t) = \exp(t\sOm)
\end{equation}
where $\sOm\in{\frak o}(3)$  is a constant matrix. We introduce the
vector $\BOm$ which is dual to $\sOm$:
$\Omega_k=-\ha\varepsilon_k^{\ ij}\Omega_{ij}$. This vector
determines the angular velocity $\Omega=|\BOm|$ and the direction
$\B n=\BOm/\Omega$ of rotation of a reference frame.

    Using the equality $\left[\sOm\B v\right]_i=(\BOm\B\times\B v)_i$ we complement
the coordinate transformation \re{2.5} by the velocity transformation:
%
\begin{equation}\label{2.6}
\dot{\B x}=\s S\B u \equiv \s S(\dot{\B z}+\sOm\B z)=\s S(\dot{\B z}+\BOm\B\times\B z)
\end{equation}
and calculate the Lagrangian in the rotating reference frame:
%
\begin{equation}\label{2.7}
\tilde L(\B z,\dot{\B z};\BOm)\equiv L(\s S\B z,\s S\B u).
\end{equation}
This function of $\B z$, $\dot{\B z}$ is rotary invariant but
with respect to the time-dependent realization of O(3):
$\B z\to \s S^{-1}(t)\s R\s S(t)\B z\equiv \s S(-t)\s R\s S(t)\B z$.
The corresponding conserved quantity is the same vector of angular momentum
$\B J$ as in eq. \re{2.4}.

Besides, the Lagrangian $\tilde L(\B z,\dot{\B z};\BOm)$
does not depend on the time $t$ explicitly, so that the conserved quantity
$\tilde E$ exists although it differs from the energy \re{2.3}:
%
\begin{equation}\label{2.8}
\tilde E = \dot{\B z}\cdot\frac{\partial\tilde L}{\partial\dot{\B
z}} - \tilde L.
\end{equation}
It is related with the integrals \re{2.3}, \re{2.4} by means of
the equality:
%
\begin{equation}\label{2.9}
\tilde E = E-\BOm\cdot\B J.
\end{equation}

\subsection{Circular orbit solutions}

Let us consider the solution of the above dynamical problem which is
static in the rotating reference frame: $\dot{\B z}=0$, $\B z=\B R$.
The Euler-Lagrange equations take the form:
%
\begin{equation}\label{2.10}
\left.\partial\tilde L/\partial\B z\right|_{\dot{\B z}=0} = 0.
\end{equation}
Taking into account the structure \re{2.7}, \re{2.2} of the Lagrangian,
the equations \re{2.10} read:
%
\begin{equation}\label{2.11}
\B R L_\alpha-\BOm\B\times(\BOm\B\times\B c)L_\gamma=0,
\end{equation}
where $L_\alpha=\partial L/\partial\alpha$ e.t.c.; in the present case
$\alpha=\B R^2$, $\beta=\B R\cdot(\BOm\B\times\B R)=0$, and
$\gamma=(\BOm\B\times\B R)^2 = \BOm^2\B R^2-(\BOm\cdot\B R)^2$.

Let us consider two cases. In the special case $\B R\parallel\BOm$ we have
$\beta=\gamma=0$ while $\alpha=R^2=|\B R|^2$ must satisfy the equation
$\alpha L_\alpha(\alpha,0,0)=0$. The solution $\B x=\s S\B R$ is
truly static:
%
\begin{equation*}
\dot{\B x}=\s S(\underbrace{\dot{\B R}}_{\vez} +
\underbrace{\BOm\B\times\B R}_{\vez})=0.
\end{equation*}
In general,
$\B R\nparallel\BOm$. Then the eq. \re{2.11} determines both
the direction of the vector $\B R$,
%
\begin{equation*}
\BOm\cdot\B R=\BOm\cdot\B x=0 \qquad \Longrightarrow \qquad \B
R\perp\BOm, \qquad \B x\perp\BOm,
\end{equation*}
as well as the relation of $R=|\B R|$ and $\Omega=|\BOm|$:
%
\begin{equation}\label{2.12}
L_\alpha(R^2,0,\Omega^2R^2)+\Omega^2L_\gamma(R^2,0,\Omega^2R^2)=0.
\end{equation}
Thus $\B R=R(\Omega)\hat{\B R}$, where $\hat{\B R}\perp\BOm$, $|\hat{\B
R}|=1$.

    The values of the integrals of motion on the circular orbit solutions
are:
%
\begin{eqnarray}
&&\B J^{(0)}=2\B R\B\times(\BOm\B\times\B R)L^{(0)}_\gamma= 2\BOm
R^2L^{(0)}_\gamma,
\label{2.13}\\
&&\tilde E^{(0)} = - L^{(0)}, \label{2.14}\\
&&E^{(0)}=2\Omega^2R^2L^{(0)}_\gamma-L^{(0)}=\Omega J^{(0)}-L^{(0)};
\label{2.15}
\end{eqnarray}
they depend on $\BOm$ only; here $L^{(0)}$, $L^{(0)}_\alpha$ e.t.c. denote values
of corresponding functions on the circular orbit solution.

\subsection{Equations of motion in the linear approximation}

Let us put
%
\begin{equation*}
\B z = \B R + \B\rho, \qquad \B u = \B v + \dot{\B\rho} +
\BOm\B\times\B\rho, \qquad\mbox{with}\quad \B v =\BOm\B\times\B R,
\end{equation*}
where $|\B\rho|\ll|\B R|$, and expand the Lagrangian \re{2.7}
in the vicinity of the extremal point $\B R$ up to quadratic (with respect
to $\B\rho$) terms. One obtains:
%
\begin{eqnarray}
\tilde L(\B z,\dot{\B z}) &=& \tilde L(\B R+\B\rho,\dot{\B\rho})
\approx\tilde L(\B R,\B0) + \underbrace{\frac{\partial\tilde L(\B
R,\B0)}{\partial\B z}}_{\vez}\cdot\B\rho +
\underbrace{\frac{\partial\tilde L(\B R,\B0)}{\partial\dot{\B
z}}\cdot\dot{\B\rho}}_{\mbox{total derivative}}
\nn\\
&&{}+\frac12\left(\frac{\partial^2\tilde L(\B R,\B0)}{\partial
z^i\partial z^j}\rho^i\rho^j+2\frac{\partial^2\tilde L(\B
R,\B0)}{\partial z^i\partial\dot
z^j}\rho^i\dot\rho^j+\frac{\partial^2\tilde L(\B
R,\B0)}{\partial\dot z^i\partial\dot z^j}\dot\rho^i\dot\rho^j
\right) \nn\\
&\equiv& L^{(0)} + L^{(2)}.
\label{2.16}
\end{eqnarray}
(the argument $\BOm$ of $\tilde L$ is omitted here). Using the
notations
%
\begin{equation}\label{2.17}
L_i\equiv\frac{\partial\tilde L(\B R,\B0)}{\partial z^i}, \qquad
L_{\hat\imath}\equiv\frac{\partial\tilde L(\B R,\B0)}{\partial\dot
z^i} \qquad \mbox{e.t.c.}
\end{equation}
we write down the second-order Lagrangian
%
\begin{equation}\label{2.18}
L^{(2)}=\ha(L_{ij}\rho^i\rho^j +2L_{i\hat\jmath}\rho^i\dot\rho^j+
L_{\hat\imath\hat\jmath}\dot\rho^i\dot\rho^j),
\end{equation}
and corresponding equations of motion:
%
\begin{equation}\label{2.19}
L_{ij}\rho^j +(L_{i\hat\jmath}-L_{j\hat\imath})\dot\rho^j-
L_{\hat\imath\hat\jmath}\ddot\rho^j=0.
\end{equation}

It is convenient to chose unit coordinate orts as follows: $\B
\epsilon_3\upuparrows\BOm$, ~~$\B \epsilon_1\upuparrows\B R$,
~~$\B \epsilon_2=\B \epsilon_3\B\times\B \epsilon_1\upuparrows \BOm\B\times\B R$, and decompose the vector
$\B\rho=\{\rho^1,\rho^2,\rho^3\}$ into coordinate components. Then
taking the rotary invariance of the Lagrangian into account (see
Appendix A) one obtains the equations of motion in the following
form:
%
\begin{eqnarray}
L_{11}\rho^1 + (L_{1\hat2}+L_{\hat2}/R)\dot\rho^2 -
L_{\hat1\hat1}\ddot\rho^1 - L_{\hat1\hat2}\ddot\rho^2 = 0,
&&\label{2.20}\\
{}-(L_{1\hat2}+L_{\hat2}/R)\dot\rho^1 - L_{\hat1\hat2}\ddot\rho^1 -
L_{\hat2\hat2}\ddot\rho^2 = 0,
&&\label{2.21}\\
-\frac{L_{\hat2}}{R\Omega}(\Omega^2\rho^3 + \ddot\rho^3) = 0,
\label{2.22}
\end{eqnarray}
The equation \re{2.22} splits out from other ones of this set; it
describes the harmonic oscillations in the direction $\B
\epsilon_3\upuparrows\BOm$ with the frequency $\Omega$. Physically, this
can be treated (in the linear approximation) as a motion of particle
along a plane orbit, the normal to which differs from $\B \epsilon_3$.
Inother words, this mode combines with a circular orbit solution
resulting a new one with the angular velocity $\tilde\BOm=\s R\BOm$
(where $\s R$ is a rotation by some small angle).
In order to avoid double counting of degrees of freedom one can
assign the constraint $\rho^3=0$.

    The equation \re{2.21} can be integrated out once:
%
\begin{equation}\label{2.23}
(L_{1\hat2}+L_{\hat2}/R)\rho^1 + L_{\hat1\hat2}\dot\rho^1 +
L_{\hat2\hat2}\dot\rho^2 =C,
\end{equation}
with the integration constant $C$. Let us show that one
can put $C=0$ without loss of generality. Indeed, if $C\ne0$, the
set of equations \re{2.20}, \re{2.23} possesses the solution
$\rho^1=\rho^1_0$, $\rho^2=\rho^2_0+\dot\rho^2_0t$ with some
constants $\rho^1_0$, $\dot\rho^2_0$ which are proportional to $C$
(the constant $\rho^2_0$ falls out the equations and is
unimportant). The variable $\rho^2$ grows beyond all bounds of
applicability of the linear approximation unless $\rho^1=0$ and
$\dot\rho^2=0$. On the other hand, the solution with $\rho^1\ne0$
or/and $\dot\rho^2\ne0$ can be treated (in the linear approximation)
as a motion of particle along the circular orbit of the radius
$\tilde R=R+\rho^1_0$ with the angular velocity $\tilde\Omega=\Omega
+ \dot\rho^2_0/R$, i.e., as some zero-order solution. Thus, it is
sufficient to put $C=0$ yielding $\rho^2=\rho^2_0$.

    Apart these two modes (in $\B\epsilon_3$ and $\B\epsilon_2$
directions) which we will refer to as
kinematic ones, the system \re{2.20}-\re{2.22} possesses the third
dynamical mode which is of physical interest. Looking for a bounded
solution of the linear set \re{2.20}-\re{2.22} we use the ansatz
%
\begin{equation}\label{2.24}
\rho^i=e^i(\omega)\mathrm e^{-\im\omega t},
\end{equation}
real part of which makes a physical sense.
Substituting this anzatz into \re{2.20}-\re{2.22} yields the set of
equations ${\cal D}(\omega)\B e(\omega)=0$ for a polarization vector
$\B e(\omega)$ with the dynamical matrix
%
\begin{equation*}
{\cal D}(\omega)=\left\|\begin{array}{ccc}
L_{11}+\omega^2L_{\hat1\hat1} & \omega^2L_{\hat1\hat2}-\im\omega \bar L_{1\hat2} & 0 \\
\omega^2L_{\hat1\hat2}+\im\omega \bar L_{1\hat2} &  \omega^2L_{\hat2\hat2} & 0 \\
0  & 0 & \frac{L_{\hat2}}{R\Omega}(\omega^2-\Omega^2)
\end{array}\right\|,
\end{equation*}
where $\bar L_{1\hat2} = L_{1\hat2} + L_{\hat2}/R$.

The secular equation:
%
\begin{equation*}
\det{\cal D}(\omega)=\frac{L_{\hat2}}{R\Omega}\omega^2(\omega^2-\Omega^2)\left\{L_{11}L_{\hat2\hat2}-\bar L_{1\hat2}^2 +
\omega^2\left[L_{\hat1\hat1}L_{\hat2\hat2}-L_{\hat1\hat2}^2\right]\right\}=0
\end{equation*}
leads to three solutions for eigenfrequencies squared corresponding
to three degrees of freedom of the system. The only one solution
corresponds to the dynamical mode:
%
\begin{equation}\label{2.25}
\omega_1^2=\frac{\bar L_{1\hat2}^2-L_{11}L_{\hat2\hat2}}{L_{\hat1\hat1}L_{\hat2\hat2}-L_{\hat1\hat2}^2}.
\end{equation}
If $\omega_1^2>0$, both values of eigenfrequencies $\pm\omega_1$ are real;
they permit us, using the ansatz \re{2.24}, to construct a physically
meaningful real and bounded solution of equations \re{2.20}--\re{2.22}.

Two other eigenfrequencies squared and corresponding eigenvectors,
%
\begin{align*}
\omega_2^2&=0,& \B e&=\{0,1,0\};\\
\omega_3^2&=\Omega^2,& \B e&=\{0,0,1\},
\end{align*}
are images of the kinematical modes mentioned above.

\subsection{Integrals of motion in the linear approximation}

Let us start from the angular momentum $\B J$. It can be presented
as follows:
%
\begin{equation}\label{2.26}
\B J=\s S\B z\B\times\s S\frac{\partial \tilde L}{\partial \dot{\B
z}}= \s S\BUp,
\end{equation}
where the vector
%
\begin{equation}\label{2.27}
\BUp\equiv\B z\B\times\frac{\partial\tilde L}{\partial \dot{\B
z}}=\frac{\partial\tilde L}{\partial\BOm}
\end{equation}
is not, in general, conserved. Its components, in the linear
approximation, can be presented as follows:
%
\begin{eqnarray}\label{2.28}
{\mathit\Upsilon}_i &=& \varepsilon_{ij}^{\ \ k}(R^j+\rho^j)(L_{\hat
k} +
L_{\hat kl}\rho^l + L_{\hat k\hat l}\dot\rho^l + \dots) \nn\\
&\approx&\varepsilon_{ij}^{\ \ k}R^jL_{\hat k} + \varepsilon_{ij}^{\
\ k}\{R^j(L_{\hat kl}\rho^l + L_{\hat k\hat l}\dot\rho^l) + L_{\hat
k}\rho^j\}\nn\\
&\equiv& {\mathit\Upsilon}_i^{(\rm0)} \qquad+ \mathit\Upsilon_i^{(\rm1)}.
\end{eqnarray}
The components $\mathit\Upsilon^{(\rm0)}_i$ and $\mathit\Upsilon^{(\rm1)}_i$
have the explicit form:
%
\begin{eqnarray}
{\mathit\Upsilon}_1^{(\rm0)} &=& {\mathit\Upsilon}_2^{(\rm0)}=0, \qquad
{\mathit\Upsilon}_3^{(\rm0)} = RL_{\hat2},
\label{2.29}\\
{\mathit\Upsilon}_1^{(\rm1)} &=& L_{\hat2}\rho^3, \qquad
{\mathit\Upsilon}_2^{(\rm1)} = -\frac1\Omega L_{\hat2}\dot\rho^3,
\label{2.30}\\
{\mathit\Upsilon}_3^{(\rm1)} &=&
\frac1R\left\{\bar L_{1\hat2}\rho^1 +
L_{\hat1\hat2}\dot\rho^1 +L_{\hat2\hat2}\dot\rho^2\right\} \equiv
\frac CR. \label{2.31}
\end{eqnarray}

    It is evidently that $\B J^{(\rm0)}=\s S\BUp^{(\rm0)}=\BUp^{(\rm0)}$.
Besides, the only kinematic modes but not the dynamical one contribute in
$\BUp^{(\rm1)}$ and thus in $\B J^{(\rm1)}=\s S\BUp^{(\rm1)}$.
It was pointed out in the previous subsection that we can put $\rho^3=0$
and $C=0$ without loss of generality. Then $\B J^{(\rm1)}=\s S\BUp^{(\rm1)}=0$,
and, in the given approximation, $\B J\approx\B J^{(\rm0)}(\BOm)$,
where the function $\B J^{(\rm0)}(\BOm)$ is defined implicitly by
\re{2.12}--\re{2.13}.

    Now we consider the energy of the system. First of all,
we calculate the correction to the zero-order term
$\tilde E^{(\rm0)}$ \re{2.14} of the integral
$\tilde E$ \re{2.8}:
%
\begin{equation}\label{2.32}
\tilde E \approx \tilde E^{(\rm0)} + \tilde E^{(\rm2)}
\equiv -L^{(\rm0)} +
\ha\{L_{\hat\imath\hat\jmath}\dot\rho^i\dot\rho^j -
L_{ij}\rho^i\rho^j\}.
\end{equation}
So that the first nontrivial correction $\tilde E^{(\rm2)}$ to
$\tilde E^{(\rm0)}$ is quadratic in $\rho^i$.
It is evidently conserved by virtue of the equations of
motion in the first-order approximation \re{2.20}-\re{2.22}.

    Further we are interested not in the integral $\tilde E$ but in the energy
%
\begin{eqnarray}\label{2.33}
E &=& \BOm\cdot\B J+\tilde E  = \dot{\B z}\cdot\frac{\partial\tilde L}{\partial\dot{\B z}} +
\BOm\cdot\frac{\partial\tilde L}{\partial\BOm}-\tilde L
\approx\BOm\cdot\B J  -L^{(\rm0)} + \tilde E^{(\rm2)}.
\end{eqnarray}
It follows from the equalities \re{2.28}--\re{2.32}:
%
\begin{eqnarray}\label{2.34}
\tilde E^{(\rm2)} &=&\ha\{L_{\hat\i\hat\j}\dot\rho^{\mathrm{i}}\dot\rho^{\,\mathrm{j}} +
L_{\mathrm{i}\mathrm{j}}\rho^{\mathrm{i}}\rho^{\,\mathrm{j}}\}
+ \frac{\Omega}{2J^{(0)}_3}\Upsilon^{(1)}_\mathrm{i}\Upsilon^{(1)\mathrm{i}}.
\end{eqnarray}
where $\mathrm{i},\mathrm{j}=1,2$. On the other hand, within the given accuracy
%
\begin{eqnarray*}
\BOm\cdot\B J&=&\Omega J_3=\Omega\sqrt{\B J^2
- J_\mathrm{i}J^\mathrm{i}}\approx\Omega\left\{J-\frac{J_\mathrm{i}J^\mathrm{i}}{2J}\right\}
\approx\Omega J-\frac{\Omega}{2J^{(0)}_3}\Upsilon^{(1)}_\mathrm{i}\Upsilon^{(1)\mathrm{i}}.
\end{eqnarray*}
Thus we obtain a useful equality
%
\begin{eqnarray}\label{2.35}
E &\approx& \Omega J  -L^{(\rm0)} + E^{(\rm2)}, \qquad\mbox{where}\quad
E^{(\rm2)}=\left.\tilde E^{(\rm2)}\right|_{\rho^3=0}
\end{eqnarray}
which holds with accuracy up to quadratic terms in $\rho$'s.

\subsection{Hamiltonian description and quantization.}

The Legendre transformation
$\dot{\B\rho}\mapsto\B\pi=\partial\tilde L/\partial\dot{\B z}$
leads to the Hamiltonian description with the Hamiltonian function
$\tilde H(\B\rho,\B\pi;\BOm)$ to be the integral of motion
$\tilde E$ \re{2.8} in terms of canonical variables $\B\rho$,
$\B\pi$.

The fixed auxiliary vector $\BOm$ was introduced to specify
the rotating reference frame and then
the circular orbit solution.
In order to generate a set of all possible circular orbit solutions
we let $\BOm$ to be a variable of angular velocity. It follows from
\re{2.27} and \re{2.33} that the Legendre transformation with respect
to both $\dot{\B z}=\dot{\B\rho}$ and $\BOm$ leads to the Hamiltonian
description with the Hamiltonian function
$H(\B\rho,\B\pi;\BUp)$ to be a conventional energy. Rotary
invariance and a Hamiltonian constraint born from the identity
\re{2.27} provides a proper balance of degrees of freedom in the
phase space enlarged with the $\BUp$ variable.

It is convenient in our case to proceed from the expression \re{2.33}.
First two terms depend on $J=|\B J|=|\BUp|$ and $\Omega$ only. Using
\re{2.12}, \re{2.13} one can express $\Omega=f(J)$. Thus, in zero-order
approximation, we have the Hamiltonian:
%
\begin{equation}\label{2.36}
H^{(0)}(J)=\left.E^{(0)}(\Omega,J)\right|_{\Omega=f(J)}.
\end{equation}
Similarly, coefficients $L_{\mathrm{i}\mathrm{j}}$, $L_{\hat\i\hat\j}$
of the quadratic form
$E^{(2)}$ in \re{2.35} turns into functions of $|\B J|$.
We note that within the Hamiltonian description the components
$J_i$ of the angular momentum $\B J$
satisfy the Poisson bracket relations (PBR):
%
\begin{equation*}
\{J_i,J_j\} = {\varepsilon_{ij}}^kJ_k\,;
\end{equation*}
non-triviality of these PBR is due to the fact that
original variables $\Omega^i$ are not velocities but quasi-velocities.

To complete the Hamiltonization one should eliminate in $E^{(2)}$
velocities $\dot\rho^\mathrm{i}$ ($\mathrm{i}=1,2$) in favour of canonical momenta
$\pi_\mathrm{i}=\partial\tilde L^{(\rm2)}/\partial\dot\rho^\mathrm{i}$
satisfying the
PBR $\{\rho^\mathrm{i},\pi_\mathrm{j}\} = \delta^\mathrm{i}_\mathrm{j}$
(others are trivial). For a
quantization purpose it is better to use normal mode complex
amplitudes $A_\alpha$ satisfying PBR:
$\{A_\alpha,A^*_\beta\}=-\im\delta_{\alpha\beta}$ ($\alpha=1,2$ in
our case). Then, using results of Subsec. II.3-4, the energy
correction $E^{(2)}$ is put in the Hamiltonian form:
%
\begin{equation}\label{2.37}
H^{(2)}=\sum\nolimits_\alpha\omega_\alpha|A_\alpha|^2=\omega_r(J)|A_r|^2.
\end{equation}
Here $\omega_r\equiv\omega_1\,({>}0)$ is the characteristic frequency of dynamical mode \re{2.25};
we redenoted the subscript $\alpha=1\to r$ to hint that this
degree of freedom corresponds to radial oscillations. The radial frequency
$\omega_r(J)$ is expressed in terms of $J$ rather than $\Omega$.

We note that the kinematic mode with the frequency $\omega_2=0$ is not oscillator-like
and must be suppressed; the corresponding solution is discussed in Subsec. II.3 after eq. \re{2.23}.
The corresponding contribution in the Hamiltonian \re{2.37} drops out automatically.

Finally we have the Hamiltonian $H=H^{(0)}+H^{(2)}$ which is ready
for quantization: variables are replaced by operators and then -- by
their eigenvalues as follows:
%
\begin{eqnarray}\label{2.38}
&&\B J\to\hat{\B J}; \qquad A_r\to\hat A_r, \quad A^*_r\to\hat A^\dag_r; \nn\\
&&J\to\sqrt{\hat{\B J}{}^2}\to\sqrt{\ell(\ell+1)}, \quad \ell=0,1,...;\nn\\
&&|A_r|^2\to\ha(\hat A_r\hat a^\dag_r+\hat a^\dag_r\hat A_r)\to n_r+\ha,
\quad n_r=0,1,...
\end{eqnarray}
It is implied, due to a perturbation procedure, the condition
$H^{(2)}\ll H^{(0)}$ which is mainly satisfied by $n_r\ll \ell$.
Then $\sqrt{\ell(\ell+1)}\approx \ell+\ha$.

General structure of the Hamiltonian \re{2.36}, \re{2.37} and its
spectrum agree completely with corresponding results
derived with the standard Hamilton-Jacobi and WKB methods.
In Appendix B the quantization method is demonstrated on the example
of a nonrelativistic particle in a power-law potential.


\section{Galilei-invariant two-particle dynamics}
\renewcommand{\theequation}{3.\arabic{equation}}
\setcounter{equation}{0}

The rotary-invariant single-particle dynamics studied in the previous section
gives us an important tool for the description of a two-particle system. Any
isolated system, anyway the non-relativistic or relativistic one, possesses
10 conserved quantities: the energy $E$, the momentum $\B P$,
the angular momentum $\B J$ and the boost $\B K$. This is consequence of
invariance under the action of a symmetry group: the Galilei group in the
non-relativistic case, and the Poincar\'e group in the relativistic case.
It is known, both in the non-relativistic and relativistic cases, that
in the rest reference frame fixed by the condition $\B P=0$ (and also $\B K=0$
in the classical, i.e., non-quantum description) the two-particle dynamics can be
reduced to an effective single-particle one with a residual symmetry group to be O(3)$\times$T,
(here T denotes the time translation group). Below we consider a non-relativistic two-particle
system and reduce it to an effective single-particle one in the ACO approximation.

\subsection{General dynamics and circular orbits}

    The Galilei-invariant two-particle Lagrangian has the following general form:
%
\begin{equation}\label{3.1}
L(\B x_1,\B x_2,\dot{\B x}_1,\dot{\B x}_2) =
\sum\limits_{a=1}^{2}\frac{m_a}{2}\dot{\B
x}_a^2 + F(\B x^2,\,\B x\cdot\dot{\B x},\,\dot{\B x}^2)
\equiv
\sum\limits_{a=1}^{2}\frac{m_a}{2}\dot{\B x_a}^2 +
F(\alpha,\beta,\gamma),
\end{equation}
where $\B x \equiv \B x_1-\B x_2$. The corresponding 10 integrals of motion are:
%
\begin{eqnarray}
E&=&\sum\limits_{a=1}^{2}\frac{m_a}{2}\dot{\B x}_a^2 +  \dot{\B
x}\cdot\frac{\partial F}{\partial\dot{\B x}} - F, \lab{3.2}\\
\B P&=&\sum\limits_{a=1}^{2}m_a\dot{\B x}_a, \lab{3.3}\\
\B J&=&\sum\limits_{a=1}^{2}m_a\B x_a\B\times\dot{\B x}_a +  \B
x\B\times\frac{\partial F}{\partial\dot{\B x}}, \lab{3.4}\\
\B K&=&\sum\limits_{a=1}^{2}m_a\B x_a - t\B P \lab{3.5}.
\end{eqnarray}

    Non-inertial variables are introduced similarly to the single-particle case:
%
\begin{equation}\label{3.6}
\B x_a=\s S\B z_a, \qquad \dot{\B x}_a=\s S\B u_a \equiv \s
S(\dot{\B z}_a+\BOm\B\times\B z_a) \equiv \s S(\dot{\B z}_a+\B v_a).
\end{equation}
In these terms the Lagrangian
%
\begin{equation}\label{3.7}
\tilde L(\B z_1,\B z_2,\dot{\B z}_1,\dot{\B z}_2;\BOm) \equiv L(\B
z,\B u_1,\B u_2)
\end{equation}
does not depend on time $t$ explicitly and thus it generates the corresponding
integral of motion:
%
\begin{equation}\label{3.8}
\tilde E = \sum\limits_{a=1}^{2}\dot{\B z}_a\cdot\frac{\partial
\tilde L}{\partial\dot{\B z}_a} - \tilde L,
\end{equation}
related to the original integrals \re{3.2}, \re{3.4} by means of
eq. \re{2.9}.

    Circular orbit solutions are determined by the conditions:
%
\begin{equation*}
\left.\partial\tilde L/\partial\B z_a\right|_{{\dot{\B
z}_1=0}\atop{\dot{\B z}_2=0}} = 0, \qquad a=1,2
\end{equation*}
which explicit form is:
%
\begin{equation}\label{3.9}
-m_a\BOm\B\times\B v_a + 2(-)^{\bar a}(\B zF_\alpha-\BOm\B\times\B
vF_\gamma) = 0, \qquad \bar a\equiv3-a.
\end{equation}
Multiplying left- and right-hand sides of these equations by $\BOm$ yields:
%
\begin{equation}\label{3.10}
2(-)^{\bar a}\BOm\B\times\B zF_\alpha = 0 \qquad\Longrightarrow\quad
\B z\perp\BOm.
\end{equation}

We note that eqs. \re{3.9} are invariant under translations along
$\BOm$, i.e., under the transformations $\B z'_a = \B z_a+\lambda\B n$
with an arbitrary $\lambda\in\Bbb R$. Indeed, it is evidently that $\B z\mapsto\B z'=\B z$,
and also
%
\begin{equation*}
\B v'_a=\BOm\B\times\B z'_a=\BOm\B\times(\B z_a+\lambda\B n)=\BOm\B\times\B
z_a=\B v_a.
\end{equation*}
Taking into account the equality \re{3.10} one finds:
%
\begin{equation*}
\BOm\cdot\B z_a=\BOm\cdot(\B z_a-(\B z_a-\B z_{\bar a}))=\BOm\cdot\B
z_{\bar a},
\end{equation*}
i.e., $z_a^\|\equiv\B n\cdot\B z_a=z_{\bar a}^\|$ but no information
for the last quantity follows from eqs. \re{3.9}. Thus one can choose
%
\begin{equation*}
\BOm\cdot\B z_a=0 \qquad\Longrightarrow\quad \B z_a\perp\BOm,
\end{equation*}
which simplifies the system \re{3.9} to the form:
%
\begin{eqnarray}\label{3.11}
\left\{m_1\Omega^2+2[F_\alpha+\Omega^2F_\gamma]\right\}\B z_1 -
2[F_\alpha+\Omega^2F_\gamma]\B z_2 &=&0, \nn\\
- 2[F_\alpha+\Omega^2F_\gamma]\B z_1 +
\left\{m_2\Omega^2+2[F_\alpha+\Omega^2F_\gamma]\right\}\B z_2&=&0,
\end{eqnarray}
where $\alpha=z^2$, $\beta=0$ and $\gamma=\Omega^2z^2$.

    The linear homogenous set of equations \re{3.11} possesses a non-trivial
solution if its determinant vanishes:
%
\begin{equation}\label{3.12}
\Omega^2\left\{m_1m_2\Omega^2+2(m_1+m_2)[F_\alpha+\Omega^2F_\gamma]\right\}=0.
\end{equation}
This is a relation between $\Omega$ and $z$.
One solution of eq.
\re{3.12} is: $\Omega=0$.
In this case the set \re{3.9} reduces to the equation
$F_\alpha\B z=0$.
If an interaction is not singular at $\B z=0$ then we have the solution
$\B z=0$.
Otherwise, the solution is determined by the equality $F_\alpha=0$
which is the condition of extremum (here, the minimum)
of the static potential of interaction.
As in the single-particle case, this solution is not suitable.

    Other roots of secular equation are determined by the condition:
%
\begin{equation}\label{3.13}
\mu\Omega^2+2[F_\alpha+\Omega^2F_\gamma]=0, \qquad \Omega\ne0
\end{equation}
(here $\mu=m_1m_2/(m_1+m_2)$ is the reduced mass) which being
combining with \re{3.11} yields the set:
%
\begin{equation}\label{3.14}
\left. {(m_1-\mu)\B z_1 + \mu\B z_2 = 0 \atop \mu\B z_1 +
(m_2-\mu)\B z_2=0}\right\}  \quad\Longrightarrow\quad \left\{
{\B
z_1 = \B R_1\equiv\frac{m_2}{m_1+m_2}\B R~~~~ \atop \B z_2 = -\B R_2 \equiv-\frac{m_1}{m_1+m_2}\B
R}\right. ,
\end{equation}
The relation of $R=|\B R|$ and $\Omega$ is, evidently, defined by \re{3.13}.

    The values of the integral of motions \re{3.2}-\re{3.5} on a the circular-orbit solutions
are as follows
%
\begin{eqnarray*}
\B P^{(0)} &=& 0, \qquad  \B K^{(0)}=0, \qquad
\tilde E^{(0)} = -\ha\mu c^2\Omega^2 - F^{(0)},
\\
\B J^{(0)} &=& \BOm
R^2(\mu+2F^{(0)}_\gamma), \qquad
E^{(0)} = R^2\Omega^2(\ha\mu+2F^{(0)}_\gamma) - F^{(0)};
\end{eqnarray*}
they obviously correspond to a rest of the system as a whole.

\subsection{The dynamics in the ACO approximation}

    Similarly to the single-particle case one puts:
\begin{eqnarray*}
\B z_a = (-)^\na\B R_a + \B\rho_a, \quad \B u_a = \B v_a + \dot{\B\rho}_a +
\BOm\B\times\B\rho_a, \qquad
\mbox{where}\quad \B v_a =(-)^\na\BOm\B\times\B R_a,
\end{eqnarray*}
and expands the Lagrangian  \re{3.7} in $\B\rho_a$,
$\dot{\B\rho}_a$. One gets $\tilde L\approx L^{(0)}+L^{(2)}$ where
%
\begin{equation}\label{3.15}
L^{(2)}=\frac12\sum\limits_{ab}(L_{ai\,bj}\rho^i_a\rho^j_b
+2L_{ai\,b\hat\jmath}\rho^i_a\dot\rho^j_b+
L_{a\hat\imath\,b\hat\jmath}\dot\rho^i_a\dot\rho^j_b),
\end{equation}
with the coefficients $L_{ai\,bj}=\left.\frac{\partial^2\tilde
L}{\partial z^i_a \partial z^j_b}\right|^{\scriptscriptstyle{(0)}}$ etc.
The corresponding equations of motion,
%
\begin{equation*}
\sum\limits_{b}\left(L_{ai\,bj}\rho^j_b
+(L_{ai\,b\hat\jmath}-L_{bj\,a\hat\imath})\dot\rho^j_b-
L_{a\hat\imath\,b\hat\jmath}\ddot\rho^j_b\right)=0,
\end{equation*}
have the following explicit form:
%
\begin{eqnarray*}
m_a(\Omega^2\rho_{ai} - \Omega_i\Omega_j\rho_a^j +
2\varepsilon_{ij}^{~~k}\Omega_k\dot\rho_a^j - \ddot\rho_{ai})
+ (-)^{\bar a}(F_{ij}\rho^j + F_{[i\hat\jmath]}\dot\rho^j -
F_{\hat\imath\hat\jmath}\ddot\rho^j) &=& 0,
\end{eqnarray*}
where $F_{ij}=\left.\frac{\partial^2\tilde F}{\partial z^i
\partial z^j}\right|^{\scriptscriptstyle{(0)}}$ etc., $F_{[i\hat\jmath]}= F_{i\hat\jmath}-F_{j\hat\imath}$
and $\B\rho=\B\rho_1-\B\rho_2$. Summing up these equations over
$a=1,2$, first with the wight 1, and then with $(-)^{\bar a}\frac{m_{\bar
a}}{m_1+m_2}$, splits these equations:
%
\begin{eqnarray}
\Omega^2\varrho_i - \Omega_i\Omega_j\varrho^j +
2\varepsilon_{ij}^{~~k}\Omega_k\dot\varrho^j - \ddot\varrho_i &=&0 \lab{3.16}\\
\left[\mu (\Omega^2\delta_{ij}-\Omega_i\Omega_j)+F_{ij}\right]
\rho^j +
[2\mu\varepsilon_{ij}^{~~k}\Omega_k+F_{[i\hat\jmath]}]\dot\rho^j
-
[\mu\delta_{ij}+F_{\hat\imath\hat\jmath}]\ddot\rho^j&=& 0,
\lab{3.17}
\end{eqnarray}
where $\B\varrho=\sum_a\frac{m_a}{m_1+m_2}\B\rho_a$ is a deviation of the center-of-mass position.

    Let us consider the equation \re{3.16}.
Choosing orts as in the 1-particle case simplifies it to the set:
%
\begin{eqnarray*}
\Omega^2\varrho^1 + 2\Omega\dot\varrho^2 - \ddot\varrho^1 &=&0 \\
\Omega^2\varrho^2 - 2\Omega\dot\varrho^1 - \ddot\varrho^2 &=&0 \\
-\ddot\varrho^3 &=& 0.
\end{eqnarray*}
In order to cut unbounded solutions off we search a solution in the form:
~~$\varrho^i = \varepsilon^i\rm e^{-\im\omega t}$,
%
%
and arrive at the secular equation:~~$\omega^2(\omega^2-\Omega^2)^2=0$.
%
%
We claim without details that eigenvectors belonging to the degenerate eigenvalue
$\omega^2=\Omega^2$ correspond to a rotation of the vector $\B\varrho$ with the frequency
$\Omega$. In the fixed (motionless) reference frame this solution is the constant
vector $\B\varepsilon\bot\BOm$. The $\omega=0$ mode possesses constant
eigenvector $\B\varepsilon\|\BOm$. All three modes can be compensated by the
translation of the origin of coordinates, i.e., by redefinition of
the center-of-mass reference frame. Thus these modes are kinematic,
and one can put $\B\varrho=0$.

    The set of equations \re{3.17} can be obtained from the set
\re{2.19} or \re{2.20}-\re{2.22} by means of formal substitution
$L\to\bar L$, where
%
\begin{equation}\label{3.18}
\bar L(\alpha,\beta,\gamma)=\ha\mu\gamma+F(\alpha,\beta,\gamma).
\end{equation}
This set leads to one dynamical mode with frequency \re{2.25}
(with the change $L\to\bar L$) and two kinematic modes in addition
to three ones described just above. Particle eigenvectors
$\B e_a$ of all the kinematic modes have the following components:
%
\begin{align}
\omega_2&=0,& \B e_1&=\{0,R_1,0\},& \B e_2&=\{0,-R_2,0\};
\label{3.19}\\
\omega_3&=\pm\Omega,& \B e_1&=\{0,0,R_1\},& \B e_2&=\{0,0,-R_2\};
\label{3.20}\\
\omega_{4,5}&=\pm\Omega,& \B e_1&=\{1,\mp\im,0\},& \B e_2&=\{1,\mp\im,0\};
\label{3.21}\\
\omega_6&=0,& \B e_1&=\{0,0,1\},& \B e_2&=\{0,0,1\},
\label{3.22}
\end{align}
where $R_a\equiv|\B R_a|$ and $\B R_a$ ($a=1,2$) are defined in
\re{3.14}.

After the kinematic modes are suppressed,
the subsequent analysis is reduced to
the single-particle case
considered in Sec. II with the effective
centre-of-mass Lagrangian
\re{3.18}, as it is in the standard treatment.


\section{Two-particle Fokker-type dynamics}
\renewcommand{\theequation}{4.\arabic{equation}}
\setcounter{equation}{0}

In this section we extend the ACO approximation method to the formalism of action integrals of Fokker type.
We start with a two-particle Fokker-type action of general form \cite{Her85}:
%
\begin{eqnarray}\lab{4.1}
I &=& \sum\limits_{a=1}^{2} \int\! dt_a L_a\left(t_a,\B x_a(t_a),\dot{\B x}_a(t_a)\right) \nn\\
&&{}+\intab dt_1 dt_2
\Phi\left(t_1,t_2,\B x_1(t_1),\B x_2(t_2),\dot{\B x}_1(t_1),\dot{\B x}_2(t_2)\right).
\end{eqnarray}
The variational problem leads to the integral-differential equations of motion:
%
\begin{equation}\lab{4.2}
\left\{\frac{\partial}{\partial\B z_a}-
\frac{d}{d t_a}\frac{\partial}{\partial\dot{\B z}_a}\right\}
\left(L_a+\Lambda_a\right)=0,\qquad a=1,2,
\end{equation}
where $\displaystyle{
\frac{d}{d t_a}\equiv
\frac{\partial}{\partial t_a} +
\dot{\B x}_a\cdot\frac{\partial}{\partial \B x_a} +
\ddot{\B x}_a\cdot\frac{\partial}{\partial \dot{\B x}_a}}$
and
%
\begin{equation}\lab{4.3}
\Lambda_1=\int_{-\infty}^{\infty} \D{}{t_2}\Phi,\qquad
\Lambda_2=\int_{-\infty}^{\infty} \D{}{t_1}\Phi.
\end{equation}

    For a physical reason we are interested mainly in
the case where the system is invariant
under transformations of the Aristotle group \cite{Sur97}
(see also \cite{Cha08}), i.e.,
time and space translations and inversions as well as space
rotations. The Aristotle group is a common subgroup of the Galilei
and Poincar\'e groups. Thus this case includes both non-relativistic
and relativistic non-local systems into consideration.

\subsection{Symmetries and conserved quantities}

Symmetry properties of the action \re{4.1} determines general structure of the functions
$L_a(t_a,\B x_a,\dot{\B x}_a)$ (farther referred to as the 1-Fokkerians) and $
\Phi(t_1,t_2,\B x_1,\B x_2,\dot{\B x}_1,\dot{\B x}_2)$ (referred to as the 2-Fokkerian) and leads to
an existence of integrals of motion studied, for non-local (i.e., Fokker-type) systems,
in \cite{Hav71,Her85}.

    The invariance under {\em time translations}, $t\to t+\lambda_0$ ($\lambda_0\in\Bbb R$),
results in the conditions:
%
\begin{align}
&X^T_0L_a\equiv\frac{\partial L_a}{\partial t_a}=0, \lab{4.4}\\
&X^T_0\Phi\equiv\sum\limits_{a=1}^{2}\frac{\partial \Phi}{\partial t_a}=0& \Longrightarrow\
\Phi(t_1,t_2,...)&=\Phi(t_1{-}t_2,...)
\equiv\Phi(\vartheta,...)
\lab{4.5}
\end{align}
and yields the {\em energy} integral of motion:
%
\begin{eqnarray}
E=\sum\limits_{a=1}^{2}\left\{\dot{\B x}_a\cdot\frac{\partial}{\partial \dot{\B x}_a}-1\right\}
\left(L_a+\Lambda_a\right) + \ints\D{}{t_1}\D{}{t_2}\frac{\partial}{\partial\vartheta}
\Phi &&\lab{4.6}\\
\mbox{where}\quad
\ints \equiv\displaystyle{\int_{-\infty}^{t_1}\int^{\infty}_{t_2}-\int^{\infty}_{t_1}\int_{-\infty}^{t_2}}.
\lab{4.7}
\end{eqnarray}

    The invariance under {\em space translations}, $\B x\to \B x+\B\lambda$ ($\B\lambda\in\Bbb R^3$),
yields the conditions:
%
\begin{align}
&\B X^TL_a\equiv\frac{\partial L_a}{\partial \B x_a}=0; \lab{4.8}\\
&\B X^T\Phi\equiv\sum\limits_{a=1}^{2}\frac{\partial \Phi}{\partial \B x_a}=0& \Longrightarrow\
\Phi(..,\B x_1,\B x_2,..)&=\Phi(..,\B x_1-\B x_2,..)
\equiv\Phi(..,\B x,..)
\lab{4.9}
\end{align}
and the conserved {\em total momentum}:
%
\begin{equation}
\B P=\sum\limits_{a=1}^{2}\frac{\partial}{\partial \dot{\B x}_a}
\left(L_a+\Lambda_a\right) - \ints\D{}{t_1}\D{}{t_2}\frac{\partial}{\partial\B x}
\Phi.
\lab{4.10}
\end{equation}

    The {\em rotary} invariance,
%
\begin{eqnarray}
&&L_a(\s{R}\dot{\B x}_a) = L_a(\dot{\B x}_a), \lab{4.11}\\
&&
\Phi(\vartheta,\s{R}\B x,\s{R}\dot{\B x}_1,\s{R}\dot{\B x}_2)=\Phi(\vartheta,\B x,\dot{\B x}_1,\dot{\B x}_2),
\lab{4.12}
\end{eqnarray}
yields the infinitesimal conditions:
%
\begin{eqnarray}
&&\B X^RL_a\equiv\dot{\B x}_a{\B\times}\frac{\partial L_a}{\partial \dot{\B x_a}}=0, \lab{4.13}\\
&&\B X^R\Phi\equiv
\B x{\B\times}\frac{\partial\Phi}{\partial \B x} +
\sum\limits_{a=1}^{2}\dot{\B x}_a{\B\times}\frac{\partial \Phi}{\partial \dot{\B x}_a}=0
\lab{4.14}
\end{eqnarray}
and results in a conservation of the {\em angular momentum} of the system:
%
\begin{multline}
\B J=
\sum\limits_{a=1}^{2}\B x_a{\B\times}\frac{\partial }{\partial \dot{\B x}_a}
\left(L_a+\Lambda_a\right)\\
-\ha\ints\D{}{t_1}\D{}{t_2}\!\!\left\{(\B x_1+\B x_2){\B\times}\frac{\partial}{\partial \B x}+
\dot{\B x}_1{\B\times}\frac{\partial}{\partial \dot{\B x}_1} -
\dot{\B x}_2{\B\times}\frac{\partial}{\partial \dot{\B x}_2}\right\}\Phi.
\lab{4.15}
\end{multline}
The consequences of discrete symmetries with respect
to space inversions and a time reversal will be considered farther.

\subsection{Fokker-type dynamics in a uniformly rotating
reference frame}

Using the non-inertial change of variables:
%
\begin{equation}\label{4.16}
\B x_a(t_a) = \s S(t_a)\B z_a(t_a)\equiv\s S_a\B z_a(t_a), \qquad \dot{\B x}_a=\s S_a\B u_a,
\end{equation}
where $\s S(t)$ and $\B u$ are defined in \re{2.5}, \re{3.6}, and symmetry properties \re{4.4}-\re{4.5},
\re{4.8}-\re{4.9}, \re{4.11}-\re{4.14} one can define "tilded" Fokkerians:
%
%
\begin{eqnarray}
L_a(\dot{\B x}_a)=L_a(\s S_a\B u_a)&=&L_a(\B u_a)\equiv \tilde L_a(\B z_a,\dot{\B z}_a;\BOm), \lab{4.17}\\
\Phi(\vartheta,\B x_1{-}\B x_2,\dot{\B x}_1,\dot{\B x}_2)
&=&\Phi(\vartheta,\s S_1\B z_1{-}\s S_2\B z_2,\s S_1\B u_1,\s S_2\B u_2)\nn\\
&=&\Phi(\vartheta,\s S_2^{\mathrm T}\s S_1\B z_1{-}\B z_2,\s S_2^{\mathrm T}\s S_1\B u_1,\B u_2)\nn\\
&=&\Phi(\vartheta,\s S(\vartheta)\B z_1{-}\B z_2,\s S(\vartheta)\B u_1,\B u_2)\nn\\
&\equiv& \tilde\Phi(\vartheta,\B z_1,\B z_2,\dot{\B z}_1,\dot{\B z}_2;\BOm).
\lab{4.18}
\end{eqnarray}

It is obviously that "tilded" Fokkerians are invariant under time translation.
Thus the corresponding integral of motion exists:
%
\begin{eqnarray}
\tilde E=\sum\limits_{a=1}^{2}\left\{\dot{\B z}_a\cdot\frac{\partial}{\partial \dot{\B z}_a}-1\right\}
\left(\tilde L_a+\tilde\Lambda_a\right) + \ints\D{}{t_1}\D{}{t_2}\frac{\partial}{\partial\vartheta}
\tilde\Phi &&\lab{4.19}
\end{eqnarray}
where the relations of $\tilde\Lambda_a$ and $\tilde\Phi$ are similar to \re{4.3}.
The equality \re{2.9} holds in the present case too which fact can be examined directly
with the use of eqs. \re{4.6}, \re{4.15} and \re{4.16}.

\subsection{Circular orbit solutions}

{\bf Proposition.} If Fokkerians are invariant with respect to the action of the Aristotle group, i.e.,
the equalities \re{4.4}, \re{4.5}, \re{4.8}, \re{4.9}, \re{4.13}, \re{4.14} hold,
the corresponding equations of motion \re{4.2} posses a circular orbit solution
with characteristics described below.
\bigskip

\noindent
{\bf Proof.}
Fokker-type equations of circular orbit motion (i.e., equations of a rest in terms of variables $\B z_a$)
have the form:
%
\begin{equation}\label{4.20}
\left.\frac{\partial}{\partial\B z_a}\left(\tilde L_a+\tilde \Lambda_a\right)\right|_{{\dot{\B
z}_1=0}\atop{\dot{\B z}_2=0}} = 0, \qquad a=1,2,
\end{equation}
%
%
\begin{eqnarray*}
\mbox{where}\quad
\left.\tilde L_a\right|_{\dot{\B z}_a=0}&=&\tilde L_a(\B z_a,0;\BOm)\equiv L^{(0)}_a(\B z_a;\BOm),\qquad\qquad a=1,2,
\\
\left.\tilde\Lambda_a\right|_{{\dot{\B
z}_1=0}\atop{\dot{\B z}_2=0}}&=&\inta\D{}{t_{\bar a}}\tilde \Phi(t_1{-}t_2,\B z_1,\B z_2,0,0;\BOm)
\qquad\quad (\bar a=3-a)
\nn\\
&=&\inta\D{}\vartheta\tilde \Phi(\vartheta,\B z_1,\B z_2,0,0;\BOm)
\equiv\Lambda^{(0)}(\B z_1,\B z_2;\BOm),
\end{eqnarray*}
and the last function is common for both values of $a=1,2$. Thus, equations of a rest
\re{4.20} can be presented
in the effective Lagrangian form:
%
\begin{equation}\label{4.21}
\frac{\partial}{\partial\B z_a}L^{(0)}\equiv
\frac{\partial}{\partial\B z_a}\left(\sum\limits_{a=1}^2 L^{(0)}_a+\Lambda^{(0)}\right)=0, \qquad a=1,2.
\end{equation}

    In order to take into account symmetric properties of Fokkerians
it is convenient to represent their general functional form as follows:
%
\begin{eqnarray}
L_a(\dot{\B x}_a) = L_a(\dot{\B x}_a^2)&\equiv&L_a(\gamma_a),\qquad\qquad a=1,2,
\lab{4.22}\\
\Phi(\vartheta,\B x,\dot{\B x}_1,\dot{\B x}_2)&=&\Phi(\vartheta,\B x^2,\B x\cdot\dot{\B x}_1,\B x\cdot\dot{\B x}_2,
\dot{\B x}_1^2,\dot{\B x}_2^2,\dot{\B x}_1\cdot\dot{\B x}_2)\nn\\
&\equiv&\Phi(\vartheta,\alpha,\beta_1,\beta_2,\gamma_1,\gamma_2,\delta).
\lab{4.23}
\end{eqnarray}
Invariance with respect to time reversal causes the property
%
\begin{equation*}
\Phi(-\vartheta,...,-\beta_1,-\beta_2,...)=\Phi(\vartheta,...,\beta_1,\beta_2,...);
\end{equation*}
the space parity is provided automatically.

    In Appendix C scalar arguments $\alpha\dots\delta$ are expressed in terms of non-inertial variables $\B z_a$
and their derivatives. Using \re{C.1}-\re{C.4} in \re{4.22}-\re{4.23} then yields "tilded" Fokkerians.

For a circular orbit problem it is sufficient to consider the static case $\dot{\B z}_a=0$. Then
the expression \re{C.1} for $\alpha$ does not change while remaining scalars
\re{C.2}-\re{C.4} simplify:
%
\begin{eqnarray*}
\beta_1^{(0)}&=&\beta_2^{(0)} =\Omega\left[\B z_1^\bot\cdot\B z_2^\bot\sin(\Omega\vartheta)-
(\B n,\B z_1,\B z_2)\cos(\Omega\vartheta)\right],
\\
\gamma_a^{(0)}&=&\B v_a^2 = \Omega^2|\B z_a^\bot|^2, \qquad
\delta^{(0)}=\Omega^2\B z_1^\bot\cdot\B z_2^\bot
\end{eqnarray*}
Upon these equalities the 2-Fokkerian takes the following general structure:
%
\begin{equation}\label{4.24}
\Phi^{(0)}=
\Phi^{(0)}(\vartheta,\B z^2,|\B z_1^\bot|^2,|\B z_2^\bot|^2,\B z_1^\bot\cdot\B z_2^\bot,(\B n,\B z_1,\B z_2);\Omega).
\end{equation}
Upon accounting the temporal reversability,
%
\begin{equation}\label{4.25}
\Phi^{(0)}(-\vartheta,...)=\Phi^{(0)}(\vartheta,...),
\end{equation}
the integrating this function over $\vartheta$ yields a general structure of $\Lambda^{(0)}$:
%
\begin{eqnarray*}
\Lambda^{(0)}&=&\inta\D{}\vartheta\Phi^{(0)}
=
\Lambda^{(0)}(\B z^2,\,|\B z_1^\bot|^2,\,|\B z_2^\bot|^2,\,\B z_1^\bot\cdot\B z_2^\bot,\,(\B n,\B z_1,\B z_2)^2;\,\Omega).
\end{eqnarray*}
Since $(\B n,\B z_1,\B z_2)=\pm|\B z_1{\B\times}\B z_2|$, so
$(\B n,\B z_1,\B z_2)^2=|\B z_1^\bot|^2|\B z_2^\bot|^2-(\B z_1^\bot\cdot\B z_2^\bot)^2$,
the final structures of the 2-Fokkerian and then of the effective Lagrangian involved in eq. \re{4.21}
is expressed via four scalar arguments:
%
\begin{eqnarray}
\Lambda^{(0)}&=&\Lambda^{(0)}(\B z^2,\,|\B z_1^\bot|^2,\,|\B z_2^\bot|^2,\,\B z_1^\bot\cdot\B z_2^\bot;\,\Omega)
\equiv
\Lambda^{(0)}(\sigma_0,\sigma_1,\sigma_2,\sigma_3;\Omega),
\lab{4.26}\\
L^{(0)}&=&L^{(0)}(\sigma_0,\sigma_1,\sigma_2,\sigma_3;\Omega)
=
\sum_{a=1}^2L_a^{(0)}(\sigma_a;\Omega)+\Lambda^{(0)}(\sigma_0\dots\sigma_3;\Omega).
\lab{4.27}
\end{eqnarray}
This form is useful for a study of circular-orbit equations \re{4.21}.

Using the notations $k_i\equiv\partial L^{(0)}/\partial\sigma_i$ ($i=0\dots3$) brings \re{4.21}
to the form:
%
\begin{equation}\label{4.28}
2(-)^{\bar a}k_0\B z+2k_a\B z^\bot_a+k_3\B z^\bot_{\bar a}=0, \qquad a=1,2.
\end{equation}
Scalar product of these equations with $\B n$ yields the condition:
%
\begin{equation*}
k_0\B n\cdot\B z=0\quad\Longrightarrow\quad\B n\cdot\B z=0\quad\Longrightarrow\quad
\B z\bot\B n
\end{equation*}
(we do not consider the solution $k_0=0$; roughly it corresponds
to a static case in a conventional meaning).

Notice that $\B z$ and thus $\sigma_0$ is translation-invariant; $\B z_a^\bot$ and thus $\sigma_a$, $\sigma_3$
are invariant under translations along $\B n$. Thus the solution of eq.\re{4.28} is specified up to
arbitrary vector along $\B n$. We fix it by means of the conditions:
%
\begin{equation*}
\B n\cdot\B z_a=0,\quad a=1,2\quad\Longrightarrow\quad
\B z_a\bot\B n, \quad\B z_a=\B z_a^\bot.
\end{equation*}
From now on both vectors $\B z_a$ are placed on a common plane and the superscript "$\bot$" can be omitted.

    The rest equations \re{4.28} take the form:
%
\begin{eqnarray}
2(k_0+k_1)\B z_1 -(2k_0-k_3)\B z_2=0,&&
\lab{4.29}\\
-(2k_0-k_3)\B z_1 +2(k_0+k_2)\B z_2=0;&&
\lab{4.30}
\end{eqnarray}
they possess non-trivial solution provided:
%
\begin{equation}\label{4.31}
4(k_0+k_1)(k_0+k_2)-(2k_0-k_3)^2=0.
\end{equation}
Then $\B z_1||\B z_2$.
Choosing orts of a moving reference frame as follows: $\B \epsilon_3=\B n\upuparrows\BOm$,
$\B \epsilon_1\upuparrows\B z_1$,
$\B \epsilon_2=\B \epsilon_3{\B\times}\B \epsilon_1$, one can recast \re{4.29}-\re{4.30} into equalities:
%
\begin{eqnarray}
&&\B z_1 =R_1\B \epsilon_1,\qquad \B z_1 =-R_2\B \epsilon_1,
\lab{4.32}\\
&&\frac{R_2}{R_1}=\frac{k_3-2k_0}{2(k_0+k_1)}=
\frac{2(k_0+k_2)}{k_3-2k_0}.
\lab{4.33}
\end{eqnarray}
By \re{4.32} we presuppose $\B z_2\uparrow\!\downarrow\B z_1$ and thus $R_1>0$, $R_2>0$,
as in the Galilei-invariant two-particle case. Otherwise $R_2<0$ but this case is rather nonphysical.

    Relations \re{4.31} and \re{4.33} form the set of equations determining
$R_1$ and $R_2$ as functions of $\Omega$ $\blacksquare$

\subsection{Integrals of motion along circular orbits}

Static character of circular orbit solutions implies that Fokkerians
$L_a^{(0)}$ and $\Lambda^{(0)}$ depend on the constant vectors \re{4.32}. Besides,
$\Phi^{(0)}$ is a function of $\vartheta=t_1-t_2$.
Thus integrals of motion can be evaluated explicitly. At that it is useful
the following "skew" integration \re{4.7} ansatz valid for arbitrary
function $f(\vartheta)$:
%
$$
\ints\D{}{t_1}\D{}{t_2}f(\vartheta)\equiv
\left[\int_{-\infty}^{t_1}\int^{\infty}_{t_2}-\int^{\infty}_{t_1}\int_{-\infty}^{t_2}\right]
\D{}{t'_1}\D{}{t'_2}f(\vartheta')
=\int_{-\infty}^{\infty}\D{}{\vartheta'}(\vartheta-\vartheta')f(\vartheta').
$$

Taking into account the time reversal \re{4.25} of $\Phi^{(0)}$ yields easily
an evaluation of the integral $\tilde E^{(0)}$:
%
\begin{align}
\tilde E^{(0)}&=-\sum\limits_{a=1}^{2}\left(L_a^{(0)}+\Lambda^{(0)}\right)
- \int_{-\infty}^{\infty}\D{}{\vartheta}\,\vartheta\frac{\partial}{\partial\vartheta}
\Phi^{(0)}\nn\\
&=-\sum\limits_{a=1}^{2}L_a^{(0)}-2\Lambda^{(0)}
+ \int_{-\infty}^{\infty}\D{}{\vartheta}
\Phi^{(0)}
=-\sum\limits_{a=1}^{2}L_a^{(0)}-\Lambda^{(0)}=-L^{(0)}\, .
\lab{4.34}
\end{align}

An evaluation of the angular momentum is cumbersome:
$\B J^{(0)}~{=}~\B n J^{(0)}$ where
%
\begin{align}
J^{(0)}=2\Omega\sum\limits_{a=1}^{2}R_a^2\left(L_a+\Lambda_a\right)_{\gamma_a}^{(0)}
&-R_1R_2\int\limits_{-\infty}^{\infty}\D{}{\vartheta}\{[(2\Phi_{\alpha}-\Omega^2\Phi_{\delta})\vartheta
+\Phi_{\beta_1}+\Phi_{\beta_2}]\sin(\Omega\vartheta)\nn\\
&+[2\Phi_{\delta}+(\Phi_{\beta_1}+\Phi_{\beta_2})\vartheta]\Omega\cos(\Omega\vartheta)\}^{(0)};
\lab{4.35}
\end{align}
the subscripts $\alpha,...,\delta$ denote derivatives, $\Phi_\alpha=\partial\Phi/\partial\alpha$ etc.

It follows from \re{4.34} and \re{4.35} the following relation:
%
\begin{equation*}
J^{(0)}=-\partial\tilde E^{(0)}/\partial\Omega=\partial L^{(0)}/\partial\Omega.
\end{equation*}
Besides, the relation \re{2.9} gives the possibility to evaluate the energy:
%
\begin{equation}\label{4.36}
E^{(0)}=\tilde E^{(0)} + \Omega J^{(0)}=-L^{(0)}+\Omega\partial L^{(0)}/\partial\Omega.
\end{equation}

The linear momentum integral vanishes: $\B P^{(0)}=0$.

\subsection{Equations of motion in oscillator approximation}

    Small perturbations $\rho_a(t_a)$ to circular orbits are introduced naturally:
\begin{equation*}
\B z_a(t_a) = (-)^\na\B R_a + \B\rho_a(t_a), \qquad
\dot{\B z}_a(t_a) = \dot{\B\rho}_a(t_a), \qquad
a=1,2
\end{equation*}
and then substituted into the action \re{4.1}. Expanding the Fokkerians up to quadratic terms in
$\rho_a(t_a)$, $\dot{\rho}_a(t_a)$ yields:
%
\begin{align}\label{4.37}
I&=\sum\limits_{a}\inta\D{}{t_a}L_a^{(0)}+\intab\D{}{t_1}\D{}{t_2}\Phi^{(0)} \nn\\
&+\sum\limits_{a}\inta\D{}{t_a}\!\Bigg\{\underbrace{\left[\frac{\partial\tilde L_a}{\partial\B z_a} +
\inta\D{}{t_{\bar a}}\frac{\partial\tilde\Phi}{\partial\B z_a}\right]^{(0)}}_{\vez}{\cdot}\B\rho_a +
\underbrace{\left[\frac{\partial\tilde L_a}{\partial\dot{\B z}_a} +
\inta\D{}{t_{\bar a}}\frac{\partial\tilde\Phi}{\partial\dot{\B z}_a}\right]^{(0)}
{\cdot}\dot{\B\rho}_a}_{\mbox{total derivative}}\!\Bigg\} \nn\\
I^{(2)}&\left\{
{\displaystyle{+\frac12}
\lefteqn{\sum\limits_{a}
\inta\D{}{t_a}\left(L_{aij}\rho_a^i\rho_a^j + 2L_{ai\hat\jmath}\rho_a^i\dot{\rho}_a^j+
L_{a\hat\imath\hat\jmath}\dot{\rho}_a^i\dot{\rho}_a^j\right)
} \atop
\displaystyle{+\frac12}
\lefteqn{\sum\limits_{a}
\sum\limits_{b}\intab\D{}{t_1}\D{}{t_2}\!\!\left(\Phi_{ai\,bj}\rho_a^i\rho_b^j
+ 2\Phi_{ai\,b\hat\jmath}\rho_a^i\dot{\rho}_b^j+
\Phi_{a\hat\imath\,b\hat\jmath}\dot{\rho}_a^i\dot{\rho}_b^j\right)
}
}\right.
\end{align}
with the coefficients $L_{aij}=\left.\frac{\partial^2\tilde L_a}{\partial
z^i_a\partial z^j_b}\right|^{\scriptscriptstyle{(0)}}$, $\Phi_{ai\,b\hat\jmath}=\left.\frac{\partial^2\tilde\Phi}{\partial
z^i_a\partial\dot{z}^j_b}\right|^{\scriptscriptstyle{(0)}}$ etc.

    The equations of motion have the form:
%
\begin{equation}\label{4.38}
\La_{aij}\rho^j_a(t)
+\La_{a[i\hat\jmath]}\dot\rho^j_a(t)-
\La_{a\hat\imath\hat\jmath}\ddot\rho^j_a(t)+
\inta\D{}{t'}\Xi_{aij}(t-t')\rho^j_{\bar a}(t')=0,
\end{equation}
where $\La_{a[i\hat\jmath]}\equiv \La_{ai\hat\jmath}-\La_{aj\hat\imath}$
($a=1,2$, $\na=3-a$),
%
\begin{align}
&\La_{aij}=L_{aij}+\Lambda_{aij},\quad \La_{ai\hat\jmath}= \dots,
\label{4.39}\\
&\Xi_{aij}(\vartheta)=\Phi_{ai\,\bar aj}(\vartheta)+(-)^{\bar a}\dot\Phi_{[ai\,\bar a\hat\jmath]}(\vartheta)-
\ddot\Phi_{{a\hat\imath}\,{\bar a\hat\jmath}}(\vartheta)
\label{4.40}
\end{align}
and $\dot\Phi(\vartheta)\equiv\partial\Phi/\partial\vartheta$.
Due to the time reversability of the dynamics, the kernel possesses the properties:
%
\begin{equation*}
\Xi_{aij}(\vartheta)=\Xi_{aji}(-\vartheta)=\Xi_{\bar aji}(\vartheta)=\Xi_{\bar aij}(-\vartheta).
\end{equation*}

    Putting $\rho_a^i(t)=e_a^i\mathrm e^{-\im\omega t}$ leads to the set of equations:
%
\begin{equation}\label{4.41}
\sum_{b=1}^2D_{ai\,bj}(\omega)e_b^j=0
\end{equation}
with the $6\times6$ dynamical matrix
%
\begin{equation}\label{4.42}
\s D(\omega)=\left\|\begin{array}{cc}
\La_{1ij}-\im\omega \La_{1[i\hat\jmath]}+\omega^2\La_{1\hat\imath\hat\jmath} & \check\Xi_{1ij}(\omega)\\
\check\Xi_{2ij}(-\omega) & \La_{2ij}-\im\omega \La_{2[i\hat\jmath]}+\omega^2\La_{2\hat\imath\hat\jmath}
\end{array}\right\|,
\end{equation}
where the off-diagonal entries
$\check\Xi_{aji}(\omega)\equiv\int\D{}{\vartheta}\Xi_{aji}(\vartheta)\mathrm e^{\im\omega\vartheta}$
possess the properties:
%
\begin{equation}\label{4.43}
\check\Xi_{aji}(\omega)=\check\Xi^\ast_{aij}(\omega),\qquad
\check\Xi_{2ij}(\omega)=\check\Xi_{1ij}(-\omega).
\end{equation}

The equation \re{4.41} determines characteristic modes of the system.
In particular, eigenfrequencies are derived from the secular equation
%
\begin{equation*}
\det{\s D}(\omega)=0.
\end{equation*}
Subsequent description of the system depends considerably on properties
of the dynamical matrix \re{4.42}.

    First of all we note that the equations of motion \re{4.38} are not
ordinary 2nd-order differential set but they form an integral-differential set
which complicates to a great extent the analysis of the dynamics.
In particular, the Cauchy problem is not appropriate and the Hamiltonization is not straightforward.
On the other hand, the set is linear which, in turns, simplifies somewhat the analysis.
The Hamiltonization scheme of a general linear system defined by a non-local action integral
is discussed in Appendix D.

\subsection{Symmetry properties of the dynamical matrix}

Similarly to the cases of ordinary single- and two-particle systems
we should separate the dynamical and kinematic modes of the dynamical matrix.
This can be done by taking the Aristotle-invariance of the system into account.
\bigskip

\noindent
{\bf Proposition.} The dynamical matrix \re{4.42} built with the Aristotle-invariant Fokkerians
admits the eigenfrequencies and eigenvectors \re{3.19}-\re{3.22}, where $R_a$ ($a=1,2$)
are determined by the equations \re{4.31}, \re{4.33}.

\bigskip

\noindent
{\bf Proof.} Multiplying l.-h.s. of equalities \re{4.8}, \re{4.9}, \re{4.13}
and \re{4.14} by $\s S_a^{-1}\equiv\s S_a^\mathrm{T}\equiv\s S^\mathrm{T}(t_a)$
and expressing them in terms of noninertial variables yields the equalities:
%
\begin{eqnarray}
\tilde X^T_{ai}\tilde L_a&=& \left\{\frac{\partial~}{\partial z_a^i} +
\varepsilon_{ik}^{\ \ l}\Omega^k\frac{\partial~}{\partial\dot z_a^l}\right\}
\tilde L_a=0,
\label{4.44}\\
\tilde X^R_{ai}\tilde L_a&=& \varepsilon_{ik}^{\ \ l}
\left\{z_a^k\frac{\partial~}{\partial z_a^l} +
\dot z_a^k\frac{\partial~}{\partial\dot z_a^l}  +
\Omega^k\varepsilon_{lm}^{\ \ n}z_a^m
\frac{\partial~}{\partial\dot z_a^n}\right\}\tilde L_a=0,
\label{4.45}\\
\tilde X^T_{ai}\tilde\Phi(\vartheta)&=&\left\{\frac{\partial~}{\partial z_a^i} +
\varepsilon_{ik}^{\ \ l}\Omega^k\frac{\partial}{\partial\dot z_a^l}\right\}
\tilde\Phi(\vartheta)\nn\\
&&{}+
[S^{(\!{-}\!1\!)^a}\!(\vartheta)]_i^{\ j}\left\{\frac{\partial}{\partial z_\na^j} +
\varepsilon_{jk}^{\ \ l}\Omega^k\frac{\partial}{\partial\dot z_\na^l}\right\}
\tilde\Phi(\vartheta)=0,
\label{4.46}\\
\tilde X^R_{ai}\tilde\Phi(\vartheta)&=& \varepsilon_{ik}^{\ \ l}
\left\{z_a^k\frac{\partial~}{\partial z_a^l} +
\dot z_a^k\frac{\partial~}{\partial\dot z_a^l}  +
\Omega^k\varepsilon_{lm}^{\ \ n}z_a^m
\frac{\partial}{\partial\dot z_a^n}\right\}\tilde\Phi(\vartheta)
+[S^{(\!{-}\!1\!)^a}\!(\vartheta)]_i^{\ j}\times
\nn\\
&&\times
\varepsilon_{jk}^{\ \ l}
\left\{z_\na^k\frac{\partial~}{\partial z_\na^l} +
\dot z_\na^k\frac{\partial~}{\partial\dot z_\na^l}  +
\Omega^k\varepsilon_{lm}^{\ \ n}z_a^m
\frac{\partial}{\partial\dot z_\na^n}\right\}\tilde\Phi(\vartheta)
=0;
\label{4.47}
\end{eqnarray}
here arguments $\B z_a$, $\dot{\B z}_a$ of $\tilde L_a$ and $\tilde\Phi$
are omitted for brevity.

Symmetry conditions \re{4.44}-\re{4.45} for 1-Fokkerians result in useful consequences:
%
\begin{align*}
\left[\tilde X^T_{ai}\tilde L_a\right]^{(0)}&=0,&
\left[\frac{\partial~}{\partial z_a^j}\tilde X^T_{ai}\tilde L_a\right]^{(0)}&=0,&
\left[\frac{\partial~}{\partial \dot z_a^j}\tilde X^T_{ai}\tilde L_a\right]^{(0)}&=0,
\\
\left[\tilde X^R_{ai}\tilde L_a\right]^{(0)}&=0,&
\left[\frac{\partial~}{\partial z_a^j}\tilde X^R_{ai}\tilde L_a\right]^{(0)}&=0,&
\left[\frac{\partial~}{\partial \dot z_a^j}\tilde X^R_{ai}\tilde L_a\right]^{(0)}&=0,
\end{align*}
where the superscript "(0)" means "on circular orbit solution", i.e., taking the conditions
$\B z_1=\B R_1$, $\B z_2=-\B R_2$, $\dot{\B z}_a=0$ into account.
These equalities impose constraints \re{E.1}-\re{E.6} for the quantities $L_{ai}$, $L_{aij}$
etc. shown in Appendix E.

For the 2-Fokkerian we are interested in the following consequences of \re{4.46}, \re{4.47}:
%
\begin{align*}
\int\limits_{-\infty}^\infty\D{}\vartheta
\left[\tilde X^T_{ai}\tilde\Phi(\vartheta)\right]^{(0)}&=0,\
\int\limits_{-\infty}^\infty\D{}\vartheta
\left[\frac{\partial~}{\partial z_a^j}\tilde X^T_{ai}\tilde\Phi(\vartheta)\right]^{(0)}=0,\
\dots,
\\
\int\limits_{-\infty}^\infty\D{}\vartheta
\left[\tilde X^R_{ai}\tilde\Phi(\vartheta)\right]^{(0)}&=0,\ \dots,\
\int\limits_{-\infty}^\infty\D{}
\vartheta\left[\frac{\partial~}{\partial \dot z_a^j}\tilde X^R_{ai}\tilde\Phi(\vartheta)\right]^{(0)}=0,
\end{align*}
which impose constraints for the quantities $\Lambda_{ai}=\check \Phi_{ai}(0)$,
$\Lambda_{aij}=\check \Phi_{aij}(0)$, \dots and $\check \Phi_{ai\na j}(\pm\Omega)$ etc.
Using the explicit form for the matrix $\s S(\vartheta)$:
%
\begin{equation*}
S_i^{\ j}(\vartheta)=\cos(\Omega\vartheta)\delta_i^j + \{1-\cos(\Omega\vartheta)\}n_in^j -
\sin(\Omega\vartheta)n^k\varepsilon_{ki}^{\ \ j},
\end{equation*}
and taking into account the equality
$\s S^{-1}(\vartheta)=\s S^\mathrm{T}(\vartheta)=\s S(-\vartheta)$ and
the fact that the Fourier-transform $\check \Phi(\omega)$ of $\tilde \Phi(\vartheta)$ and
its derivatives $\check \Phi_{ai}(\omega)$ etc.
are pair-vise functions of $\omega$, one can arrive at the equalities
\re{E.7}-\re{E.12} in Appendix E. They lead together with the equations
\re{E.1}-\re{E.6} and the equations of a rest $\La_{ai}=0$
to the following linear relations for elements of the dynamical matrix
$\s D(0)$ and $\s D(\Omega)$:
%
\begin{eqnarray}
&&\La_{ai3}+\Lambda_{ai\,\na3}=0,
\label{4.48}\\
&&R_a\La_{ai2}-R_\na\Lambda_{ai\,\na2}=0,
\label{4.49}\\
&&
R_a(\La_{ai3}+\Omega^2\La_{a\hat\imath\hat3})
-R_\na(\check\Phi_{ai\,\na3}(\Omega)+\Omega^2\check\Phi_{{a\hat\imath}\,{\na\hat3}}(\Omega))=0,
\label{4.50}\\
&&R_a\La_{a[i\hat3]}-R_\na\check\Phi_{[ai\,{\na\hat3}]}(\Omega)=0,
\qquad a={1,2},\quad i={1,2,3},\qquad
\label{4.51}\\
&&\La_{ai\mathrm{j}}-\Omega\varepsilon_{3\mathrm{j}}^{\ \ k}\La_{a[i\hat k]}+\Omega^2\La_{a\hat\imath\hat\j}
+ \check{\Phi}_{ai\,\na \mathrm{j}}(\Omega)\nn\\
&&\hspace{5ex}{}-\Omega\varepsilon_{3\mathrm{j}}^{\ \ k}\check{\Phi}_{[ai\,{\na\hat k}]}(\Omega)
+\Omega^2\check{\Phi}_{{a\hat\imath}\,{\na\hat\j}}(\Omega) =0,\ \ \mathrm{j}={1,2.}
\label{4.52}
\end{eqnarray}
The relations \re{4.48} provide the existence of the eigenfrequency
and the eigenvector \re{3.22}, the relations \re{4.49} -- of \re{3.19},
the relations \re{4.50}-\re{4.51} -- of \re{3.20}, and
the relations \re{4.52} -- of \re{3.21} $\blacksquare$
\bigskip

Thanks to this proposition, the kinematic modes separate from physical modes,
eigenfrequencies of which can be found from the equation:
%
\begin{equation}
\frac{\det{\s D}(\omega)}{\omega^4(\omega^2-\Omega^2)^3}=0.
\label{4.53}
\end{equation}

    Since $\s D(\omega)$ is a 6$\times$6 matrix, entries of which are not,
in general, polynomial, the secular equation \re{4.53} is rather complicated.
It can be simplified somewhat due to the following
\skip 0.5cm

\noindent
{\bf Proposition.} The following identities hold:
%
\begin{equation*}
D_{a\mathrm{j}\,b3}(\omega)=D_{a3\,b\mathrm{j}}(\omega)=0 \quad\mbox{for any}\ a,b=1,2\ \mbox{and}\ \mathrm{j}=1,2.
\end{equation*}

\bigskip

\noindent
{\bf Proof} can be completed directly using the representation \re{4.22}, \re{4.23} of 1-
and 2-Fokkerians and examining the following equalities:
%
\begin{eqnarray}
\left.\frac{\partial\alpha}{\partial z_a^3}\right|^{(0)}=\dots
=\left.\frac{\partial\delta}{\partial z_a^3}\right|^{(0)}=0,\qquad
\left.\frac{\partial\alpha}{\partial\dot z_a^3}\right|^{(0)}=\dots
=\left.\frac{\partial\delta}{\partial\dot z_a^3}\right|^{(0)}=0,\nn\\
\left.\frac{\partial^2\alpha}{\partial z_a^\mathrm{j}\partial z_b^3}\right|^{(0)}=\dots=0, \qquad
\dots=\left.\frac{\partial^2\delta}{\partial\dot z_a^\mathrm{j}\partial\dot z_b^3}\right|^{(0)}=0,
\qquad \mathrm{j}=1,2,
\nn
\end{eqnarray}
where $\alpha,\dots\delta$ are defined by \re{C.1}-\re{C.4} in Appendix C, and the superscript
"(0)" denotes the value of a marked quantity on the circular orbit solution $\blacksquare$
\bigskip

Thanks to this proposition, $\det\s D(\omega)$ splits into two factors:
%
\begin{equation*}
\det\s D(\omega)=\det\s D^\bot(\omega)\cdot\det\s D^\|(\omega),
\end{equation*}
%
%
\begin{equation*}
\mbox{where}\quad
\s D^\bot(\omega)=\left\|D_{a\mathrm{i}\,b\mathrm{j}}(\omega)\right\| \quad(\mathrm{i,j}=1,2),
\quad \s D^\|(\omega)=\left\|D_{a3\,b3}(\omega)\right\|.
\end{equation*}
The 2$\times$2 submatrix $\s D^\|(\omega)$ possesses two kinematic modes \re{3.20} and \re{3.22}
while the 4$\times$4 submatrix $\s D^\bot(\omega)$ -- another three kinematic modes \re{3.19}, \re{3.21}
and one the dynamical mode. The frequency of the latter can be determined fr0m the reduced secular equation:
%
\begin{equation}
\frac{\det{\s D^\bot}(\omega)}{\omega^2(\omega^2-\Omega^2)^2}=0.
\label{4.54}
\end{equation}

The secular equations \re{4.53} or \re{4.54} simplify in the case of equal particles.

\subsection{The dynamics of the equal particle system}

The equal particles system is defined by Fokkerians of the following properties:
%
\begin{eqnarray}
L_a(\dot{\B x}_a)&=&L(\dot{\B x}_a),
\lab{4.55}\\
\Phi(\vartheta, \B x_1, \B x_2, \dot{\B x}_1, \dot{\B x}_2)&=&
\Phi(-\vartheta, \B x_2, \B x_1, \dot{\B x}_2, \dot{\B x}_1).
\lab{4.56}
\end{eqnarray}
%

\noindent
{\bf Proposition.} Along with a plausible assumption the equations of motion \re{4.20}
for equal particles possess a circular motion solution of the form:
%
\begin{equation}
\B z_1=\B R,\qquad\B z_2=-\B R \qquad(\mbox{and}\quad
\dot{\B z}_1=\dot{\B z}_2=0)
\label{4.57}
\end{equation}
with characteristics described below.

\bigskip

\noindent
{\bf Proof.} It follows from \re{4.57} and \re{4.29}, \re{4.30} the equality: $k_1=k_2$
which turns into identity provided \re{4.57} holds.
Then r.h.s of the equation \re{4.31} for $|\B R|=R(\Omega)$ factorizes:
%
\begin{equation}
(4k_0+2k_1-k_3)=0 \quad\mbox{or/and}\quad (2k_1+k_3)=0.
\label{4.58}
\end{equation}
Although solutions of both the equations \re{4.58} must be examined,
the only first equation seems to make a physical sense since it includes the
derivative $k_0=\partial L^{(0)}/\partial\B z^2$  which is intuitively related to
a fource of interparticle interaction $\blacksquare$
\bigskip

\noindent
{\bf Proposition.} If the equal particle system is invariant under
the space inversions the entries of the dynamical
matrix satisfy the equalities:
%
\begin{equation}
D_{2i\,2j}(\omega)=D_{1i\,1j}(\omega), \qquad
D_{2i\,1j}(\omega)=D_{1i\,2j}(\omega)
\label{4.59}
\end{equation}
%

\noindent
{\bf Proof.} If 1-Fokkerian is invariant under space inversions,
%
\begin{equation*}
\tilde L(-\B z,-\dot{\B z})=\tilde L(\B z,\dot{\B z}),
\end{equation*}
then its second-order derivatives are invariant too. Thus we have
%
\begin{equation}\label{4.60}
\textstyle{ L_{2ij}=\frac{\partial^2\tilde L}{\partial z^i\partial
z^j}(-\B R,0) =\frac{\partial^2\tilde L}{\partial z^i\partial
z^j}(\B R,0)=L_{1ij} }
\end{equation}
and, similarly,
%
\begin{equation}\label{4.61}
L_{2i\hat\jmath}=L_{1i\hat\jmath},\qquad L_{2\hat\imath\hat\jmath}=L_{1\hat\imath\hat\jmath}.
\end{equation}

The 2-Fokkerian for equal particles can be presented in the form
%
\begin{equation*}
\tilde\Phi(\vartheta,\B z_1,\B z_2,\dot{\B z}_1,\dot{\B z}_2)=
\ha F(\vartheta,\B z_1,\B z_2,\dot{\B z}_1,\dot{\B z}_2) +
\ha F(-\vartheta,\B z_2,\B z_1,\dot{\B z}_2,\dot{\B z}_1);
\end{equation*}
here $F(\vartheta,\B x,\B y,\B u,\B v)$ is some function
(for example, the $\Phi(\vartheta,\B x,\B y,\B u,\B v)$ itself) which is
inversion-invariant,
%
\begin{equation*}
F(\vartheta,-\B x,-\B y,-\B u,-\B v) =
F(\vartheta,\B x,\B y,\B u,\B v),
\end{equation*}
and thus its second-order derivatives are so. We have the equality
%
\begin{eqnarray*}
\Phi_{2i\,2j}(\vartheta)&=&
\textstyle{
\frac{\partial^2F}{2\partial y^i\partial y^j}(\vartheta,\B R,-\B R,0,0)
+\frac{\partial^2F}{2\partial x^i\partial x^j}(-\vartheta,-\B R,\B R,0,0)
}\nn\\
&=&
\textstyle{
\frac{\partial^2F}{2\partial y^i\partial y^j}(\vartheta,-\B R,\B R,0,0)
+\frac{\partial^2F}{2\partial x^i\partial x^j}(-\vartheta,\B R,\B R,0,0)
}
=\Phi_{1i\,1j}(-\vartheta)
\end{eqnarray*}
and similar equalities for other derivatives and their Fourier-transforms:
%
\begin{equation}\label{4.62}
\check\Phi_{2i\,2j}(\omega)=\check\Phi_{1i\,1j}(-\omega),\dots,
\check\Phi_{2\hat\imath\,1\hat\jmath}(\omega)=\check\Phi_{1\hat\imath\,2\hat\jmath}(-\omega).
\end{equation}
Then using the definitions \re{4.39}, \re{4.40}, \re{4.42} and the properties \re{4.43}, \re{4.60},
\re{4.61} and \re{4.62} results in the equalities \re{4.59} $\blacksquare$
\bigskip

Thanks to equalities \re{4.59} the set of equations \re{4.41} splits into to subsets:
%
\begin{eqnarray}
{\frak D}_{ij}(\omega)\varepsilon^j=0\ &\mbox{with}&\
{\frak D}_{ij}(\omega)\equiv
\La_{1ij}-\im\omega \La_{1[i\hat\jmath]}+\omega^2\La_{1\hat\imath\hat\jmath} + \check\Xi_{1ij}(\omega)
\nn\\
&& =D_{1i\,1j}(\omega)+D_{1i\,2j}(\omega), \ \varepsilon^j\equiv\ha(e_1^j{+}e_2^j);
\lab{4.63}\\
{\cal D}_{ij}(\omega)e^j=0,\ &\mbox{with}&\
{\cal D}_{ij}(\omega)\equiv
\La_{1ij}-\im\omega \La_{1[i\hat\jmath]}+\omega^2\La_{1\hat\imath\hat\jmath} - \check\Xi_{1ij}(\omega)
\nn\\
&& = D_{1i\,1j}(\omega)-D_{1i\,2j}(\omega), \quad e^j\equiv e_1^j-e_2^j. \lab{4.64}
\end{eqnarray}

It is easy to verify that the subset \re{4.63} possesses three kinematic eigenfrequencies
and eigenvectors \re{3.21}, \re{3.22} while the subset \re{4.64} -- another two kinematic modes
\re{3.19}, \re{3.20} and one the dynamical mode.
The frequency of the latter can be determined from the reduced secular equation:
%
\begin{equation*}
\frac{\det{{\cal D}^\bot}(\omega)}{\omega^2}=0 \quad\mbox{where}\
{\cal D}^\bot(\omega)\equiv\|{\cal D}_{\mathrm{ij}}(\omega)\| \quad (\mathrm{i,j}=1,2)
\end{equation*}
is the reduced dynamical matrix. In the next subsection we discuss possible solutions of this equation
as well as of more general equations \re{4.53} or \re{4.54}.

\subsection{Predictive treatment of the Fokker-type system.}

Let us call a particle system as {\em predictive} if it possesses three
degrees of freedom per particle. For example it is the Galilei-invariant
Lagrangian two-particle system considered in Sec. III.
Consequently, in ACO approximation
the corresponding dynamical matrix admits 6 modes with
real frequencies, 5 of which have a kinematic origin
(they are 0 or $\pm\Omega$), and only one mode characterizes a specific dynamics
(it corresponds to radial interparticle oscillations).

The same kinematic modes arise in Aristotle-invariant
Fokker-type 2-particle system and thus, in a Lagrangian
system (as a particular case when 2-Fokkerian includes
$\delta(\vartheta)$) as well as in the Poincar\'e-invariant
Fokker-type system (as a particular case with extra
Lorentz-invariance).

In contrast to predictive systems, the Fokker-type dynamical system
may possess an infinite number of degrees of freedom, due to a time nonlocality
of equations of motion.
In ACO approximation the quadratic term $I^{(2)}$ of the action \re{4.37}
(i.e., 3rd and 4th strings) is of the Fokker-type too.
Consequently, the dynamical matrix of such systems
is not polynomial and may possess an infinitely large number of modes with
real or/and complex frequencies. It has been shown in Subsec. IV.6 and IV.7 how
kinematic modes can be separated out. The question arises: how one could understand
the case when a number of remaining modes is more then 1 ?

One point of view is that extra degrees of freedom are inherent to Fokker-type system
considered literally as a physical model. If complex frequencies are present, the system
is unstable and thus it cannot form bound states, at least in the ACO approximation.
It has been pointed out in Appendix D that some modes with real frequencies may
contribute negatively in the energy what is another kind of instability \cite{P-U50}.
In these cases the model should be adjudged as physically inconsistent.

One can adhere another point of view. We considered two-particle Fokker-type action integral
where 1-Fokkerians correspond to a free-particle system while 2-Fokkerian
describes particle interaction. A system of two free particles possesses 6 degrees of freedom.
The same is true for the interacting system describing within the predictive Lagrangian dynamics.
If one endows a particle interpretation of the Fokker-type system, the latter should
possess 6 physical degrees of freedom. Thus extra degrees of freedom should be considered as a
mathematical artifact of the Fokker formalism or specific model, and finally should be separated out
the physical dynamics of the system. Similar situation arises when considering the Lorentz-Dirac equation
\cite{YaT12}. It possesses an extra solution describing exponentially accelerating particle
even if external forces vanish. This solution is commonly discarded as nonphysical.

The question is how to separate physical degrees of freedom out of nonphysical ones ?
In our case, how to recognize the only dynamical mode of radial excitations ?
A kind of selection rule can be suggested if 1) there exist some parameter of nonlocality
$\tau$ such that:
%
\begin{equation*}
\Phi_\tau(\vartheta, \B x_1,...,\dot{\B x}_2) \mathop{\longrightarrow}\limits_{\tau\to0}
\delta(\vartheta)\Lambda(\B x_1,...,\dot{\B x}_2),
\end{equation*}
i.e., in the limit $\tau\to0$ the system turns into a predictive Lagrangian system,
and 2) if this predictive system admits circular orbit solution.
As a consequence, all modes of the dynamical matrix reduce to 5 kinematic
and 1 dynamical ones while every extra mode disappears. It is possible,
for example, if corresponding frequency
$|\omega|\mathop{\longrightarrow}\limits_{\tau\to0}0$ (then this mode does not contribute
in the Hamiltonian; see \re{2.37}, \re{D.18} and \re{D.22})
or $|\omega|\mathop{\longrightarrow}\limits_{\tau\to0}\infty$
(then it never can be excited).

    If there is no an explicit parameter of nonlocality, it sometimes can be defined dynamically, as
a function of the angular velocity $\Omega$ or of the angular momentum $J$. This is possible because
$\Omega$ is a parameter with respect to the action $I^{(2)}$ in \re{4.37} as well as the angular
momentum $J$ or the quantum number $\ell$ are parameters as to the classical or quantum hamiltonian $H^{(2)}$
\re{2.37}. For example, in the Fokker action formulation of electrodynamics \cite{A-B72} one can put
$\tau\sim v = R\Omega$, where $v$ is a particle speed along a circular orbit of the radius
$R$. In the domain $v\ll1$ there exists only one mode with the frequency which can be identified as $\omega_r(J)$.
By continuity this mode can be recognized and selected in the essentially relativistic domain $v\lesssim1$, while
other (extra) modes should be discarded as nonphysical.

Similarly, one can treat other relativistic systems.

After kinematic modes are suppressed
and nonphysical modes are discarded
the subsequent treatment of the two-particle Fokker-type system
reduces to the effective single-particle case
with the effective Hamiltonian
%
\begin{equation}\label{4.65}
H_{\rm eff}(J,|A_r|)=H^{(0)}(J)+H^{(2)}(J,|A_r|)=
H^{(0)}(J)+ \omega_r(J)|A_r|^2,
\end{equation}
where $J=|\B J|=\sqrt{\B J^2}$ and $|A_r|=\sqrt{A_r^*A_r}$.

At this point one should refer to a symmetry of the original Fokker-type
system which is Galilei-invariant in a non-relativistic case or
Poincar\'e-invariant in a relativistic case. In both cases
the effective Hamiltonian is understood as the energy of the system in
the center-of-mass (CM) reference frame. It is a function of $\B J$
which is meant as the intrinsic angular momentum of the system, and $A_r$
which is the amplitude of interparticle radial oscillations.
In order to have a complete Hamiltonian description of the system one must
introduce variables (or operators) characterizing the state of the system
as a whole, for example, the total momentum $\B P$ and the canonically conjugated
CM position $\B Q$. Together with $\B J$, $A_r$ and the function
\re{4.65} these variable unambiguously determine a canonical realization
of a symmetry (Galileo or Poincar\'e) group, i.e., the complete Hamiltonian
description of the system.

For example, the total energy of a non-relativistic system is
%
\begin{equation}\label{4.66}
H=\ha(m_1{+}m_2)^{-1}\B P^2+H_{\rm eff}(J,|A_r|).
\end{equation}
In the relativistic case:
%
\begin{equation}\label{4.67}
H=\sqrt{M^2+\B P^2},  \qquad\mbox{where}\quad M=H_{\rm eff}(J,|A_r|),
\end{equation}
i.e., the effective Hamiltonian coincides with the total mass $M$ of the system,
while the total Hamiltonian \re{4.67} and other generators of the
Poincar\'e group are determined in terms of $M$, $\B J$, $\B P$ and $\B Q$ via
the Bakamjian-Thomas (BT) or equivalent model \cite{B-T53,Duv89}.
The quantization of BT model is well elaborated \cite{Sok78,Pol89}. But the spectrum of the
mass operator can be obtained from \re{4.65} by means of the substitution \re{2.38}.


\section{Conclusion}
\renewcommand{\theequation}{5.\arabic{equation}}
\setcounter{equation}{0}

In this paper we proposed a quantization scheme for a two-particle
Fokker-type system on ACO approximation. For generality it is
considered a system which is invariant under the Aristotle group,
the common subgroup of the Galileo and the Poincar\'e groups. In
such a way both non-relativistic as well as relativistic systems are
included into consideration. And only at a very final stage the
scheme refers to a genuine symmetry of the system.

It has been proven that the Aristotle-invariant two-particle system
admits a planar circular motion of particles with arbitrary (in
principle) angular velocity $\BOm$ provided some rather general
conditions hold. This is done by the usage of a non-inertial uniformly
rotating reference frame in which circular orbits are search as
static (equilibrium) solutions of equations of motion. Radii $R_a$
of particle orbits are stated as certain implicit functions of
$\Omega=|\BOm|$. Then small perturbations of particle motion around
equilibrium point are considered. They correspond to almost circular
particle orbits \cite{A-B72} as referred by the inertial observer.

The action principle of Fokker type for perturbed motion is derived.
It leads to a set of linear homogeneous integral-differential
equation. General properties of this set has been studied.

It is shown by means of a group-theoretical analysis that a possibly
wide variety of characteristic modes of this set includes a number
of modes which are unessential and must be suppressed in order to
avoid a double count of degrees of freedom. Also, there is a one
mode corresponding to radial inter-particle oscillations. Our
attitude is that, namely, this mode is physically meaningful. All
other characteristic modes (if exist) are unstable or physically
unacceptable and must be suppressed. The corresponding selection
rule is suggested. The reduced dynamical system is equivalent to
some effective single-body problem. It is put by means of the Llosa
procedure \cite{L-V94} into the Hamiltonian formulation which then
is expanded to a two-body problem in accordance to a symmetry of the
origrnal Fokker-type system. For example, if the original
Fokker-type system is Poincar\'e-invariant, the final Hamiltonian
description is formulated within the Bakamjian-Thomas (BT) or
equivalent model \cite{B-T53,Duv89} on the 12-dimensional phase space
$\Bbb P$. In other words, as a dynamical system, this BT model is
a predictive subsystem of the original Fokker-type system. A subsequent
quantization is straightforward.

The presented scheme for quantization of
Fokker-type models may appear useful in the nuclear and hadronic
physics. The Fokker-type model of Regge trajectories
will be presented in a forthcoming paper \cite{Duv12b}.

\section*{Acknowledgment}
The author is grateful to
V. Tretyak and Yu.~Yaremko for helpful discussion of this work.


\section*{Appendix A. Rotary invariance properties of a single particle Lagrangian}
\renewcommand{\theequation}{A.\arabic{equation}}
\setcounter{equation}{0}

    We use the rotary invariance property \re{2.1} of
the Lagrangian \re{2.2} in the infinitesimal form:
%
\begin{equation}
X^R_i L = \varepsilon_{ik}^{\ \ l}\left(x^k\frac{\partial
}{\partial x^l} + \dot x^k\frac{\partial }{\partial\dot
x^l}\right)L=0. \lab{A.1}
\end{equation}
Applying the infinitesimal operators $\tilde X^R_i\equiv S^j_{\ i}X^R_j$
to the Lagrangian \re{2.7} we express the identities \re{A.1} in
terms of the variables $\B z$, $\dot{\B z}$:
%
\begin{equation}
\tilde X^R_i\tilde L= \varepsilon_{ik}^{\ \
l}\left(z^k\frac{\partial\tilde L}{\partial z^l} + \dot
z^k\frac{\partial\tilde L}{\partial\dot z^l}  +
\Omega^k\varepsilon_{lm}^{\ \ n}z^m\frac{\partial\tilde
L}{\partial\dot z^n}\right)=0. \label{A.2}
\end{equation}
Then the equation \re{2.10}, the identities \re{A.2} and their
differential consequences
%
\begin{equation*}
\frac{\partial}{\partial z^j}\tilde X^R_i \tilde L = 0, \qquad
\frac{\partial}{\partial \dot z^j}\tilde X^R_i \tilde L = 0,
\end{equation*}
taken on circular orbit, result in the homogeneous linear set of equations:
%
\begin{eqnarray*}
&& L_i=0, \qquad
\varepsilon_{ik}^{\ \ l}
\Omega^k\varepsilon_{lm}^{\ \ n}z^m L_{\hat n} = z_i\Omega^n L_{\hat
n} = 0,
\\
&&\varepsilon_{ik}^{\ \ l}\left(z^k L_{lj} +
\Omega^k(\varepsilon_{lj}^{\ \ n} L_{\hat n} + \varepsilon_{lm}^{\ \
n}z^m L_{\hat nj})\right)
= \delta_{ij}\Omega^nL_{\hat n} -
L_{\hat\imath}\Omega^j + \varepsilon_{ik}^{\ \ l}z^kL_{lj}
-z_i\Omega^n L_{\hat nj}=0,
\\
&& \varepsilon_{ij}^{\ \ l}L_{\hat l} + \varepsilon_{ik}^{\ \
l}\left(z_kL_{l\hat\jmath} + \Omega^k\varepsilon_{lm}^{\ \ n}z^m
L_{\hat n\hat\jmath}\right)=
\varepsilon_{ij}^{\ \ l}L_{\hat l} +
\varepsilon_{ik}^{\ \ l}z_kL_{l\hat\jmath} + z_i\Omega^nL_{\hat
n\hat\jmath} = 0,
\end{eqnarray*}
where $\B z=\B R$. This set permits us to express a one part
of coefficients \re{2.17} $L_i$, $L_{ij}$ etc. via another part of them.
Choosing unit coordinate orts as described before eq. \re{2.20} we have:
%
$$\|L_{ij}\|=\|L_{ji}\|=\left\|\begin{array}{ccc}
L_{11}&0&0\\
0&0&0\\
0&0&-\frac{\Omega}{R}L_{\hat2} \end{array}\right\|,
\qquad
\|L_{\hat\imath\hat\jmath}\|=\|L_{\hat\jmath\hat\imath}\|=\left\|\begin{array}{ccc}
L_{\hat1\hat1}&L_{\hat1\hat2}&0\\
L_{\hat1\hat2}&L_{\hat2\hat2}&0\\
0&0&\frac{1}{R\Omega}L_{\hat2} \end{array}\right\|,
$$
$$\|L_{i\hat\jmath}\|=\|L_{\hat\jmath i}\|=\left\|\begin{array}{ccc}
L_{1\hat1}&L_{1\hat2}&0\\
-\frac1RL_{\hat2}&\frac1RL_{\hat1}&0\\
0&0&\frac{1}{R}L_{\hat1} \end{array}\right\|
\quad\mbox{so that}\quad
\|L_{i\hat\jmath}-L_{\hat\jmath i}\|=\left\|\begin{array}{ccc}
0&L_{1\hat2}+\frac1RL_{\hat2}&0\\
-L_{1\hat2}-\frac1RL_{\hat2}&0&0\\
0&0&0 \end{array}\right\|.
$$
It follows from these formulae the equations \re{2.20}-\re{2.22}.

\section*{Appendix B. Nonrelativistic particle in a power-law
potential}
\renewcommand{\theequation}{B.\arabic{equation}}
\setcounter{equation}{0}

Let us consider the following Lagrangian of a non-relativistic
particle:
%
\begin{equation}\label{B.1}
L=\frac m2 \dot{\B x}^2 - a|\B x|^\nu
= \frac m2\gamma -a\alpha^{\nu/2}, \qquad a\nu>0.
\end{equation}
Eqs. \re{2.12}, \re{2.13} and \re{2.15} in this case lead to the following circular orbit
equation and integrals of motion:
%
\begin{eqnarray}
\Omega^2&=&\nu\frac{a}{m}R^{\nu-2},
\label{B.2}\\
J&=&mR^2\Omega,
\label{B.3}\\
E^{(0)}&=&\ha mR^2\Omega^2+aR^\nu.
\label{B.4}
\end{eqnarray}
Combining
\re{B.2} with \re{B.3} and \re{B.4}
yields the
functions which determine the circular orbit Hamiltonian:
%
\begin{eqnarray}
\Omega&=&\left[(\nu a)^2\frac{J^{\nu-2}}{m^\nu}\right]^{\frac1{\nu+2}},
\label{B.5}\\
H^{(0)}&=&\left(\frac\nu2+1\right)\left[a^2\left(\frac{J^2}{\nu m}\right)^\nu\,\right]^{\frac1{\nu+2}}.
\label{B.6}
\end{eqnarray}

The radial frequency \re{2.25} the present case,
$\omega^2=\nu(\nu{+}2)\frac am R^{\nu-2}$, simplifies with the use of \re{B.2}:
%
\begin{equation}\label{B.7}
\omega_r=\sqrt{\nu+2}\,\Omega.
\end{equation}
It follows from this that the circular motion is unstable at $\nu\le-2$.

Gathering \re{B.6}, \re{2.37}, \re{B.7}, \re{B.5} all together and
using the quantization rules \re{2.38} one obtains the energy spectrum:
%
\begin{equation}\label{B.8}
E=\left(\frac\nu2+1\right)\left[a^2\left(\frac{\ell(\ell+1)}{\nu m}\right)^\nu\,\right]^{\frac1{\nu+2}}
\left\{1+\frac{\nu}{\sqrt{\nu+2}}\frac{2n_r+1}{\sqrt{\ell(\ell+1)}}\right\}.
\end{equation}
It coincides exactly with that formula derived in \cite{LSG91}
by solving the Schr\"odinger equation in the oscillator approximation.

In the cases of Coulomb and oscillator potentials we have:
%
\begin{eqnarray}
\nu&=&-1\,:\nn\\
E&=&-\frac{a^2m}{2\ell(\ell+1)}
\left\{1-\frac{2n_r+1}{\sqrt{\ell(\ell+1)}}\right\}
=-\frac{a^2m}{2(\ell+n_r+1)^2}+O(\ell^{-4});
\nn\\
\nu&=&2\,:\nn\\
E&=&\sqrt{\frac{2a}{m}\ell(\ell+1)}
\left\{1+\frac{2n_r+1}{\sqrt{\ell(\ell+1)}}\right\}
=\sqrt{\frac{2a}{m}}\left\{\ell+2n_r+\frac32\right\}+O(\ell^{-1}).
\nn
\end{eqnarray}

\section*{Appendix C. Scalar arguments of Fokkerians}
\renewcommand{\theequation}{C.\arabic{equation}}
\setcounter{equation}{0}
\vspace{-5ex}

%
\begin{eqnarray}
\alpha&=&\B x^2=(\s S_1\B z_1-\s S_2\B z_2)^2=(\s S\B z_1-\B z_2)^2\nn\\
&=&\B z^2+2\left\{\B z_1^\bot\cdot\B z_2^\bot[1-\cos(\Omega\vartheta)]
-(\B n,\B z_1,\B z_2)\sin(\Omega\vartheta)\right\},
\lab{C.1}\\
\beta_a&=&\dot{\B x}_a\cdot\B x=
(-)^{\bar a}(\dot{\B z}_a+\B v_a)\cdot(\B z_a-\s S_a^{\mathrm T}\s S_{\bar a}\B z_{\bar a})\nn\\
&=&(-)^{\bar a}(\dot{\B z}_a+\B v_a)\cdot\left\{\B z_a-\B z_{\bar a}\cos(\Omega\vartheta) -
\B n(\B n\cdot\B z_{\bar a})[1-\cos(\Omega\vartheta)]\right.\nn\\
&&\left.{}-(-)^a\B n{\B\times}\B z_{\bar a}\sin(\Omega\vartheta)\right\}, \qquad a=1,2;\quad\bar a=3-a,
\lab{C.2}\\
\gamma_a&=&\dot{\B x}_a^2=[\s S_a(\dot{\B z}_a+\B v_a)]^2 = \dot{\B z}_a^2+2\dot{\B z}_a\cdot\B v_a+\B v_a^2,
\lab{C.3}\\
\delta&=&\dot{\B x}_1\cdot\dot{\B x}_2=[\s S_1(\dot{\B z}_1+\B v_1)]\cdot[\s S_2(\dot{\B z}_2+\B v_2)]=
[\s S(\dot{\B z}_1+\B v_1)]\cdot(\dot{\B z}_2+\B v_2)\nn\\
&=&(\dot{\B z}_1+\B v_1)\cdot(\dot{\B z}_2+\B v_2)\cos(\Omega\vartheta) +
(\B n\cdot\dot{\B z}_1)(\B n\cdot(\dot{\B z}_2)[1-\cos(\Omega\vartheta)]\nn\\
&&{}+ (\B n,\dot{\B z}_1+\B v_1,\dot{\B z}_2+\B
v_2)\sin(\Omega\vartheta); \lab{C.4}
\end{eqnarray}
here $\B v_a\equiv\BOm{\B\times}\B z_a$ is a vector product of $\BOm$ and $\B z_a$;
$(\B n,\B z_1,\B z_2)=\B n\cdot(\B z_1{\B\times}\B z_2)$;
$\s S_a=\exp(t_a\s\Omega)$ ($a=1,2$); $\s S=\s S_1\s S_2^{\mathrm T}=\exp(\vartheta\s\Omega)$

\section*{Appendix D. Hamiltonization of a system of nonlocal oscillators}
\renewcommand{\theequation}{D.\arabic{equation}}
\setcounter{equation}{0}

    Quadratic terms of the action \re{4.37} can be presented in the following simple form:
%
\begin{equation}\lab{D.1}
I^{(2)} = \ha\sum\limits_{kl}\intab dtdt'
\rho^k(t)D_{kl}(t-t')\rho^l(t'),
\end{equation}
where the matrix kernel $\s D(t-t')=||D_{kl}(t-t')||$
(here the multindeces are used: $k,l=(a,i),(b,j)$ etc.)
is invariant with respect to time translations and
a time reversal:\\ $\s D^{\rm
T}(t'-t)=\s D(t-t')$.
The time-nonlocal linear equations of motion:
%
\begin{equation}\lab{D.2}
\sum\limits_l\inta dt' D_{kl}(t-t')\rho^l(t')=0
\end{equation}
admit a fundamental set of solutions of the form:
$\rho^k(t)=e^k(\omega)\mathrm{e}^{-\im\omega t}$. Their substitution
into the equations \re{D.2} yields the set of algebraic equations:
%
\begin{equation}\lab{D.3}
\sum\limits_lD_{kl}(\omega)e^l(\omega)=0,
\end{equation}
which constitute the eigenvector-eigenvalue problem for the
polarization vectors $e^k(\omega)$ and characteristic frequencies
$\omega$. The latters are determined from the secular equation $\det
\sf D(\omega)=0$ in terms of the dynamical matrix: $ {\s
D}(\omega)=\int\! dt\, \s D(t){\rm e}^{\im\omega t}$. (Here the
Fourier-image ${\s D}(\omega)$ is denoted by the same symbol as the
prototype ${\s D}(t)$ but with different argument which might not
lead to confusion). Due to non-locality of the problem
\re{D.2} matrix elements $ D_{kl}(\omega)$ are non-polynomial, in
general, functions of $\omega$. Consequently, solutions of the
secular equation form, in general, an infinitely large set of
complex and/or real characteristic frequencies. Thanks to symmetry
properties of the dynamical matrix:
%
\begin{equation}\lab{D.4}
 {\s D}^{\rm T}(\omega)= {\s D}(-\omega),~~~ {\s D}^\dag(\omega)
= {\s D}(\omega^*)
\end{equation}
this set consists of quadruplets
$\{\pm\omega_\alpha,\pm\omega^*_\alpha,\ \alpha=1,2,\dots\}$ (and/or
duplets if $\omega_\alpha\in\Bbb R$), and a general real solution of
the equations \re{D.2} is:
%
\begin{equation}\lab{D.5}
\rho^k(t)=\sum\limits_\alpha\left\{A_\alpha
e^k_\alpha\ {\rm e}^{-\im\omega_\alpha t}+
A^*_\alpha \cc{e}^k_\alpha\ {\rm
e}^{\im\omega^*_\alpha t}
+
B_\alpha f^k_\alpha\ {\rm e}^{-\im\omega^*_\alpha t}+
\smash{B^*_\alpha
\cc{f}^k_\alpha}\ {\rm e}^{\im\omega_\alpha t}
\right\},
\end{equation}
where $e^k_\alpha\equiv e^k(\omega_\alpha)$ and $f^k_\alpha\equiv e^k(\omega^*_\alpha)$.

    Arbitrary complex variables $A_\alpha$ and $B_\alpha$
parameterize a phase space ${\Bbb P}^{\Bbb C}$ of the system
(infinitely-dimensional, in general) which may include both the
physical as well as non-physical degrees of freedom. Complex
frequencies cause necessarily difficulties in a physical
interpretation of the system which we discuss below (also see
\cite{P-U50}). Thus, from the whole variety of frequencies we choose
only real ones: $\omega^*_\alpha=\omega_\alpha$. The general
solution \re{D.5} reduces in this case to the following one:
%
\begin{equation}\lab{D.6}
\rho^k(t)=\sum\limits_\alpha\left\{A_\alpha
e^k_\alpha\ {\rm e}^{-\im\omega_\alpha t}+
A^*_\alpha \cc{e}^k_\alpha\ {\rm
e}^{\im\omega_\alpha t}\right\},
\end{equation}
where summation spreads over those $\alpha$ for which
$\mbox{Im}~\omega_\alpha=0$. This is implied in the next subsection too.

\subsection*{D.1. Hamiltonian description: real frequencies}

    A current problem at this point is to construct the Hamiltonian
description for the variational principle \re{D.1} on the phase
subspace ${\Bbb P}^{\Bbb R}\subset{\Bbb P}^{\Bbb C} $ of solutions
\re{D.6} parameterized by complex variables $A_\alpha$. An
appropriate guideline which we adopt for this purpose is the
Hamiltonian formalism for nonlocal Lagrangians proposed by Llosa and
coauthors \cite{JJLM87,L-V94,JJLM89}.

    Let us define the time-nonlocal lagrangian:
%
\begin{equation}\lab{D.7}
L(t) = \ha\sum\limits_{kl}\inta dt'
\rho^k(t)D_{kl}(t-t')\rho^l(t'),
\end{equation}
in term of which the action \re{D.1} takes a usual form $I^{(2)} =
\inta dtL(t)$, and the functional derivative:
%
\begin{equation}\lab{D.8}
E_k(t,t';[\rho]) \equiv \frac{\delta L(t)}{\delta\rho^k(t')}=
\ha\sum_l\rho^l(t)D_{lk}(t-t').
\end{equation}
Then, following the Ref. \cite{L-V94}, the Hamiltonian structure on the phase space
of the time-nonlocal system is defined by the symplectic form, i.e.,
the closed differential 2-form:
%
\begin{eqnarray}\lab{D.9}
\Omega&=&\sum_{kl}\inta dt\inta ds\inta du\chi(t,s)
\frac{\delta E_k(-s,t;[\rho])}{\delta\rho^l(u)}\delta\rho^l(u)\wedge\delta\rho^k(t),\\
\mbox{where}&&\chi(t,s)\equiv\ha(\mathrm{sgn}t+\mathrm{sgn}s)=\theta(t)\theta(s)-\theta(-t)\theta(-s),\nn
\end{eqnarray}
$\delta\rho^k(t)$ denotes a functional differential of $\rho^k(t)$, and "$\wedge$" denotes
the wedge product. In turns, an explicit calculation of
the symplectic form determines PBR of phase variables (in our case - of $A_\alpha$'s).

It is evidently that the symplectic form \re{D.9} is exact, i.e.,
$\Omega=\delta\Theta$, where
%
\begin{equation}\lab{D.10}
\Theta=\sum_{k}\inta dt\inta ds\chi(t,s)E_k(-s,t;[\rho])\delta\rho^k(t)
\end{equation}
is a 1-form, so called the Liouville form, defined up to arbitrary exact
1-form (i.e., a total differential). A calculation of $\Theta$ (rather than $\Omega$)
is more simple and convenient for our purpose.

    The dynamics of a time-nonlocal system in a phase space is determined
by the Hamiltonian \cite{L-V94}:
%
\begin{equation}\lab{D.11}
H=\sum_{k}\inta dt\inta ds\,\chi(t,s)E_k(-s,t;[\rho])\dot\rho^k(t)-L(t)|_{t=0}.
\end{equation}

    Upon integration in eqs. \re{D.10} and \re{D.11} the following formula is useful:
%
\begin{equation*}
\inta dt\inta ds\,\chi(t,s)E(-s,t)f(t)=\int\limits_{-\infty}^\infty\! ds\int\limits_0^s\! dt\, E(t-s,t)f(t).
\end{equation*}

    Let us calculate the Liouville form $\Theta$. Using \re{D.6} and \re{D.8} in
\re{D.10} yields
%
\begin{multline}\lab{D.12}
\Theta=\ha\sum_{kl}\int\limits_{-\infty}^\infty \!\!ds\int\limits_0^s \!\!dt\, D_{kl}(s)\rho^l(t-s)\delta\rho^k(t)\\
=\ha\sum_{kl}\sum_{\alpha\beta}\int\limits_{-\infty}^\infty\!\!ds\, D_{kl}(s)\int\limits_0^s\!\!dt
\left(A_\alpha e^l_\alpha\mathrm{e}^{-\im\omega_\alpha(t-s)}
+A^*_\alpha \cc{e}^l_\alpha\mathrm{e}^{\im\omega_\alpha(t-s)}\right)\times\\
\times\left(\mathrm{e}^{-\im\omega_\beta t}e^k_\beta \delta A_\beta +
\mathrm{e}^{\im\omega_\beta t}\cc{e}^k_\beta \delta A^*_\beta\right)
\end{multline}
which then is convenient to present in the matrix form:
%
\begin{eqnarray}\lab{D.13}
\Theta&=&\ha\sum_{kl}\sum_{\alpha\beta}\int\limits_{-\infty}^\infty\!\!ds\, D_{kl}(s)
[
\begin{array}{ccc}
A_\alpha e^l_\alpha\mathrm{e}^{\im\omega_\alpha s},&&
A^*_\alpha \cc{e}^l_\alpha\mathrm{e}^{-\im\omega_\alpha s}
\end{array}
]\times\hspace{15ex}\nn
\\
&&\hspace{21ex}
\times\int\limits_0^s\!\!dt
\left[
\begin{array}{ccc}
\vphantom{\cc{e}^k_\beta}\mathrm{e}^{-\im(\omega_\alpha+\omega_\beta)t}
&& \mathrm{e}^{-\im(\omega_\alpha-\omega_\beta)t}\\
\vphantom{\cc{e}^k_\beta}\mathrm{e}^{\im(\omega_\alpha-\omega_\beta)t}
&& \mathrm{e}^{\im(\omega_\alpha+\omega_\beta)t}
\end{array}
\right]
\left[
\begin{array}{c}
e^k_\beta\delta A_\beta\\
\cc{e}^k_\beta\delta A^*_\beta
\end{array}
\right]\nn\\
&=&\hi\sum_{kl}\sum_{\alpha\beta}
[
\begin{array}{ccc}
A_\alpha e^l_\alpha,&&
A^*_\alpha \cc{e}^l_\alpha
\end{array}
]\times\nn
\\
&&\hspace{10ex}
\times\int\limits_{-\infty}^\infty\!\!ds\, D_{kl}(s)
\left[
\begin{array}{ccc}
-\frac{\mathrm{e}^{\im\omega_\alpha s}-\mathrm{e}^{-\im\omega_\beta s}}{\omega_\alpha+\omega_\beta}&
-\frac{\mathrm{e}^{\im\omega_\alpha s}-\mathrm{e}^{\im\omega_\beta s}}{\omega_\alpha-\omega_\beta}\\
\frac{\mathrm{e}^{-\im\omega_\alpha s}-\mathrm{e}^{-\im\omega_\beta s}}{\omega_\alpha-\omega_\beta}&
\frac{\mathrm{e}^{-\im\omega_\alpha s}-\mathrm{e}^{\im\omega_\beta s}}{\omega_\alpha+\omega_\beta}
\end{array}
\right]
\left[
\begin{array}{c}
e^k_\beta\delta A_\beta\\
\cc{e}^k_\beta\delta A^*_\beta
\end{array}
\right]\nn\\
&=&\hi\sum_{kl}\sum_{\alpha\beta}
[
\begin{array}{ccc}
A_\alpha e^l_\alpha,&&
A^*_\alpha \cc{e}^l_\alpha
\end{array}
]\times\nn
\\
&&\hspace{10ex}
\times
\left[
\begin{array}{ccc}
-\frac{ D_{lk}(\omega_\alpha)-D_{lk}(-\omega_\beta)}{\omega_\alpha+\omega_\beta}&
-\frac{ D_{lk}(\omega_\alpha)- D_{lk}(\omega_\beta)}{\omega_\alpha-\omega_\beta}\\
\frac{ D_{lk}(-\omega_\alpha)- D_{lk}(-\omega_\beta)}{\omega_\alpha-\omega_\beta}&
\frac{ D_{lk}(-\omega_\alpha)- D_{lk}(\omega_\beta)}{\omega_\alpha+\omega_\beta}
\end{array}
\right]
\left[
\begin{array}{c}
e^k_\beta\delta A_\beta\\
\cc{e}^k_\beta\delta A^*_\beta
\end{array}
\right]\!\!.
\end{eqnarray}
Due to properties \re{D.3} and \re{D.4} all items of the sum
\re{D.13} vanish except those terms which both correspond to
$\alpha=\beta$ and include anti-diagonal entries of the square
matrix in the last line of \re{D.13}. Residuary terms possess
uncertainty 0/0 which can be eliminated by a limit transition:
%
\begin{eqnarray}\lab{D.14}
\Theta&=&\hi\sum_{kl}\sum_{\alpha}\lim\limits_{\lambda\to\omega}
\left(
A^*_\alpha \cc{e}^l_\alpha
\frac{ D_{lk}(\lambda_\alpha)- D_{lk}(\omega_\alpha)}{\lambda_\alpha-\omega_\alpha}
e^k_\alpha\delta A_\alpha
-
A_\alpha e^l_\alpha
\frac{ D_{lk}(-\lambda_\alpha)- D_{lk}(-\omega_\alpha)}{\lambda_\alpha-\omega_\alpha}
\cc{e}^k_\alpha\delta A^*_\alpha
\right)\nn
\\
&=&\hi\sum_{\alpha}\Delta_\alpha(A^*_\alpha\delta A_\alpha -A_\alpha\delta A^*_\alpha),
\end{eqnarray}
where
%
\begin{equation}\lab{D.15}
\Delta_\alpha\equiv\sum_{kl}\cc{e}^k_\alpha
\frac{d D_{kl}(\omega_\alpha)}{d\omega}
e^l_\alpha=\Delta_\alpha^*.
\end{equation}

    In order to calculate the Hamiltonian we first note that the Lagrangian \re{D.7}
equals to zero by virtue of the equations of motion \re{D.2}, thus
the last term of \re{D.11} vanishes. The residuary sum in \re{D.11}
can be obtained from the Liouville form \re{D.10} by means of
formal substitution $\delta A_\alpha\to -\im\omega_\alpha A_\alpha$.
Thus one gets:
%
\begin{equation}\lab{D.16}
H=\sum_{\alpha}\Delta_\alpha\omega_\alpha|A_\alpha|^2.
\end{equation}

    If $\Delta_\alpha>0$ one can redefine polarization vectors
$e^k_\alpha\to\tilde e^k_\alpha=\Delta_\alpha^{-1/2}e^k_\alpha$ in
\re{D.6} so that the Liouville form and the Hamiltonian simplify:
%
\begin{eqnarray}
\Theta&=&\hi\sum_{\alpha}(A^*_\alpha\delta A_\alpha -A_\alpha\delta A^*_\alpha),
\lab{D.17}\\
H&=&\sum_{\alpha}\omega_\alpha|A_\alpha|^2. \lab{D.18}
\end{eqnarray}
Then one gets from \re{D.17} the symplectic form:
%
\begin{equation}\lab{D.19}
\Omega=\delta\Theta=\im\sum_{\alpha}\delta A^*_\alpha\wedge\delta A_\alpha
\end{equation}
which generates the following PBR:
$\{A_\alpha,A_\beta\}=\{A^*_\alpha,A^*_\beta\}=0$,
$\{A_\alpha,A^*_\beta\}=-\im\delta_{\alpha\beta}$. Upon quantization
$A_\alpha\to\hat A_\alpha$, $A^*_\alpha\to\hat A_\alpha^\dag$ one
obtains standard annihilation and creation operators: $[\hat
A_\alpha,\hat A_\beta^\dag]=\delta_{\alpha\beta}$; the Hamiltonian
\re{D.18} takes the standard oscillator form and leads immediately
to the discrete spectrum $E=\sum_\alpha\omega_\alpha(n_\alpha+\ha)$,
$n_\alpha=0,1,\dots$. We will refer to variables $A_\alpha$,
$A^*_\alpha$, and the form \re{D.19} as canonical ones.

Let us now $\{\alpha\}=\{\alpha'\}\cup\{\alpha''\}$ such that
$\Delta_{\alpha'}>0$ and $\Delta_{\alpha''}<0$. Redefining the
polarization vectors $e^k_\alpha\to\tilde
e^k_\alpha=|\Delta_\alpha|^{-1/2}e^k_\alpha$ and then the canonical
variables $A_{\alpha''}\to C_{\alpha''}=A^*_{\alpha''}$ reduce the
Liouville form \re{D.14} to the canonical one:
%
\begin{eqnarray*}
\hspace{-3ex}
\Theta&=&\hi\sum_{\alpha'}(A^*_{\alpha'}\delta A_{\alpha'} {-} A_{\alpha'}\delta A^*_{\alpha'}) -
\hi\sum_{\alpha''}(A^*_{\alpha''}\delta A_{\alpha''} {-} A_{\alpha''}\delta A^*_{\alpha''})\nn\\
&=&\hi\sum_{\alpha'}(A^*_{\alpha'}\delta A_{\alpha'} {-} A_{\alpha'}\delta A^*_{\alpha'}) +
\hi\sum_{\alpha''}(C^*_{\alpha''}\delta C_{\alpha''} {-} C_{\alpha''}\delta C^*_{\alpha''}).
\qquad\quad
\end{eqnarray*}
The Hamiltonian in this case,
%
\begin{equation*}
H=\sum_{\alpha'}\omega_{\alpha'}|A_{\alpha'}|^2 - \sum_{\alpha''}\omega_{\alpha''}|C_{\alpha''}|^2,
\end{equation*}
is not positively defined which
feature
is
characteristic of
higher derivative and time-nonlocal systems.
\cite{P-U50}.

\subsection*{D.2. Hamiltonian description: complex frequencies}

If $\mbox{Im}~\omega_\alpha\ne0$, the corresponding terms in
\re{D.5} are unbounded which fact contradicts the taken
approximation of small $\rho^k$. One can chose a dumping solution by
putting $B_\alpha=0$ in \re{D.5}. This case however cannot be
turned into the Hamiltonian formalism, using the scheme by Llosa
et.al. \cite{L-V94}. To see this we consider briefly a general case
of complex frequencies.

Substituting the solution \re{D.5} into \re{D.10} and
accomplishing similar to eqs. \re{D.12}-\re{D.14} (but more
cumbersome) calculations yields the Liouville form:
%
\begin{equation}\lab{D.20}
\Theta=\hi\sum_{\alpha}\left\{\Delta_\alpha(B^*_\alpha\delta A_\alpha {-} A_\alpha\delta B^*_\alpha)
+\Delta_\alpha^*(A^*_\alpha\delta B_\alpha {-} B_\alpha\delta A^*_\alpha)\right\},
\end{equation}
with complex (contrary to \re{D.15}) coefficients
%
\begin{equation*}
\Delta_\alpha\equiv\sum_{kl}\cc{f}^k_\alpha
\frac{dD_{kl}(\omega_\alpha)}{d\omega}
e^l_\alpha\ne\Delta_\alpha^*.
\end{equation*}
Then the redefinition $e^k_\alpha\to\tilde
e^k_\alpha=\Delta_\alpha^{-1/2}e^k_\alpha$, $f^k_\alpha\to\tilde
f^k_\alpha=\cc{\Delta}_\alpha^{-1/2}f^k_\alpha$ in \re{D.5} reduce
\re{D.20} to the form:
%
\begin{equation}\lab{D.21}
\Theta=\hi\sum_{\alpha}\left(B^*_\alpha\delta A_\alpha {-} A_\alpha\delta B^*_\alpha
+A^*_\alpha\delta B_\alpha {-} B_\alpha\delta A^*_\alpha\right).
\end{equation}
Similarly one obtains the Hamiltonian:
%
\begin{equation}\lab{D.22}
H=\sum_\alpha\left(\omega_\alpha B^*_\alpha A_\alpha
+\omega_\alpha^*A^*_\alpha B_\alpha\right).
\end{equation}

    The Liouville form \re{D.21} is not split in variables $A_\alpha$ and $B_\alpha$ which
thus are not canonical nor appropriate for quantization. Properties of the Hamiltonian
\re{D.22} in these variables are obscured. Thus we change variables into canonical ones.
For a brevity we consider only one mode corresponding to some quadruplet
of characteristic frequencies. Thus hereinafter the indices $\alpha$
and summation over $\alpha$ are omitted.

A choice of canonical variables is not unique. One may choose for one a complex
variables $a$, $b$ as follows:
%
\begin{equation*}
A=(a-b^*)/\sqrt{2}, \qquad B=(a+b^*)/\sqrt{2},
\end{equation*}
in terms of which the Liouville form indeed becomes separated and canonical:
%
\begin{eqnarray*}
\Theta&=&\hi(B^*\delta A - A\delta B^* + A^*\delta B - B\delta A^*)
=\hi(a^*\delta a - a\delta a^* + b^*\delta b - b\delta b^*).
\end{eqnarray*}
On the contrary, the Hamiltonian does not split in $a$ and $b$ modes,
%
\begin{equation*}
H=\ha(\omega B^*A + \omega^*A^*B)
=\ha\left\{\mathrm{Re}\,\omega(a^*a - b^*b) + \im\mathrm{Im}\,\omega(ba - b^*a^*)\right\}.
\end{equation*}
In this case, real canonical variables are more appropriate for a quantization.
If one chooses:
%
\begin{equation*}
A=(-P+\im p)/\sqrt{2}, \qquad B=(q+\im Q)/\sqrt{2},
\end{equation*}
the Liouville form standardizes:~~
$\Theta\by{md}p\delta q + P\delta Q$~~
(the notation ``md'' means ``up to a total differential'') while~~
$H=-\mathrm{Re}\,\omega(qP-Qp)- \im\mathrm{Im}\,\omega(qp+QP)$.~~
In this form the Hamiltonian has been quantized in \cite{P-U50}
and its spectrum is shown to be continuous and unbounded both
from below and from above. A physical meaning of such a system is
doubtful.

\section*{Appendix E. Aristotle-invariance properties of the dynamical
matrix\vspace{-2ex}}
\renewcommand{\theequation}{E.\arabic{equation}}
\setcounter{equation}{0}
%
\begin{eqnarray}
&&L_{ai}+\varepsilon_{ik}^{\ \ l}\Omega^k L_{a\hat l}=0,
\lab{E.1}\\
&&L_{aij}+\varepsilon_{ik}^{\ \ l}\Omega^k L_{a\hat l j}=0,
\lab{E.2}\\
&&L_{ai\hat\jmath}+\varepsilon_{ik}^{\ \ l}\Omega^k L_{a\hat l\hat\jmath}=0,
\lab{E.3}\\
&&\varepsilon_{ik}^{\ \ l}\{z_a^k L_{al} + \Omega^k\varepsilon_{lm}^{\ \ n}z_a^m L_{a\hat n}\}=0,
\lab{E.4}\\
&&\varepsilon_{ik}^{\ \ l}\{z_a^k L_{alj} + \Omega^k\varepsilon_{lm}^{\ \ n}z_a^m L_{a\hat nj}\}
+ \varepsilon_{ij}^{\ \ l} L_{al} + \varepsilon_{ik}^{\ \ l}\varepsilon_{lj}^{\ \ n}\Omega^k L_{a\hat n}
=0,\qquad
\lab{E.5}\\
&&\varepsilon_{ik}^{\ \ l}\{z_a^k L_{al\hat\jmath} + \Omega^k\varepsilon_{lm}^{\ \ n}z_a^m L_{a\hat n\hat\jmath}\}
+ \varepsilon_{ij}^{\ \ l} L_{a\hat l}=0,
\lab{E.6}\\
&&\Lambda_{ai}+\varepsilon_{ik}^{\ \ l}\Omega^k\Lambda_{a\hat l} +
n_in^m\{\Lambda_{\na n}+\varepsilon_{nk}^{\ \ l}\Omega^k\Lambda_{\na\hat l}\} \nn\\
&&\hspace{10ex}{}+ \Pr\nolimits_i^n\{\check\Phi_{\na n}(\Omega)+\varepsilon_{nk}^{\ \ l}\Omega^k\check\Phi_{\na\hat l}(\Omega)\}
=0,
\lab{E.7}\\
&&\Lambda_{aij}+\varepsilon_{ik}^{\ \ l}\Omega^k\Lambda_{a\hat lj} +
n_in^n\{\Lambda_{\na n\,aj}+\varepsilon_{nk}^{\ \ l}\Omega^k\Lambda_{{\na\hat l}\,aj}\} \nn \\
&&\hspace{10ex}{}+ \Pr\nolimits_i^n\{\check\Phi_{\na n\,aj}(\Omega)
+\varepsilon_{nk}^{\ \ l}\Omega^k\check\Phi_{{\na\hat l}\,aj}(\Omega)\}
=0,
\lab{E.8}\\
&&\Lambda_{ai\hat\jmath}+\varepsilon_{ik}^{\ \ l}\Omega^k\Lambda_{a\hat l\hat\jmath} +
n_in^n\{\Lambda_{{\na n}\,{a\hat\jmath}}
+\varepsilon_{nk}^{\ \ l}\Omega^k\Lambda_{{\na\hat l}\,{a\hat\jmath}}\} \nn\\
&&\hspace{10ex}{}+ \Pr\nolimits_i^n\{\check\Phi_{\na n\,{a\hat\jmath}}(\Omega)
+\varepsilon_{nk}^{\ \ l}\Omega^k\check\Phi_{{\na\hat l}\,{a\hat\jmath}}(\Omega)\}
=0,
\lab{E.9}\\
&&\varepsilon_{ik}^{\ \ l}\{z_a^k \Lambda_{al} + \Omega^k\varepsilon_{lm}^{\ \ n}z_a^m \Lambda_{a\hat n}\}  \nn\\
&&\hspace{5ex}{}+ n_in^q\varepsilon_{qk}^{\ \ l}\{z_a^k \Lambda_{\na l}
+ \Omega^k\varepsilon_{lm}^{\ \ n}z_a^m \Lambda_{\na\hat n}\}
\nn\\
&&\hspace{10ex}{}+ \Pr\nolimits_i^q
\varepsilon_{qk}^{\ \ l}\{z_a^k \check\Phi_{\na l}(\Omega)
+ \Omega^k\varepsilon_{lm}^{\ \ n}z_a^m \check\Phi_{\na\hat n}(\Omega)\}
=0,
\lab{E.10}\\
&&\varepsilon_{ik}^{\ \ l}\{z_a^k \Lambda_{alj} + \Omega^k\varepsilon_{lm}^{\ \ n}z_a^m \Lambda_{a\hat nj}\}  \nn\\
&&\hspace{5ex}{}+ n_in^q\varepsilon_{qk}^{\ \ l}\{z_a^k \Lambda_{\na l\,aj} +
\Omega^k\varepsilon_{lm}^{\ \ n}z_a^m \Lambda_{{\na\hat n}\,aj}\}
\nn\\
&&\hspace{10ex}{}+ \Pr\nolimits_i^q
\varepsilon_{qk}^{\ \ l}\{z_a^k \check\Phi_{\na l\,aj}(\Omega)
+ \Omega^k\varepsilon_{lm}^{\ \ n}z_a^m \check\Phi_{{\na\hat n}\,aj}(\Omega)\} \nn\\
&&\hspace{15ex}{} + \varepsilon_{ij}^{\ \ l} \Lambda_{al}
+ \varepsilon_{ik}^{\ \ l}\varepsilon_{lj}^{\ \ n}\Omega^k \Lambda_{a\hat n}
=0,
\lab{E.11}\\
&&\varepsilon_{ik}^{\ \ l}\{z_a^k \Lambda_{al\hat\jmath}
+ \Omega^k\varepsilon_{lm}^{\ \ n}z_a^m \Lambda_{{a\hat n}{\hat\jmath}}\}  \nn\\
&&\hspace{10ex}{}+ n_in^q\varepsilon_{qk}^{\ \ l}\{z_a^k \Lambda_{\na l\,{a\hat\jmath}} +
\Omega^k\varepsilon_{lm}^{\ \ n}z_a^m \Lambda_{{\na\hat n}\,{a\hat\jmath}}\}
\nn\\
&&\hspace{1ex}{}+ \Pr\nolimits_i^q \varepsilon_{qk}^{\ \ l}\{z_a^k
\check\Phi_{\na l\,{a\hat\jmath}}(\Omega) +
\Omega^k\varepsilon_{lm}^{\ \ n}z_a^m \check\Phi_{{\na\hat
n}\,{a\hat\jmath}}(\Omega)\} + \varepsilon_{ij}^{\ \ l}
\Lambda_{a\hat l} =0, \lab{E.12}
\end{eqnarray}
where $\B z_1=\B R_1$, $\B z_2=-\B R_2$ and $\Pr\nolimits_i^k\equiv\delta_i^k-n_in^k$.

\newpage


\end{document}